# Good Concatenated Code Ensembles for the Binary Erasure Channel

Alexandre Graell i Amat, *Member, IEEE,* and Eirik Rosnes, *Member, IEEE,*

*Abstract*—In this work, we give good concatenated code ensembles for the binary erasure channel (BEC). In particular, we consider repeat multiple-accumulate (RMA) code ensembles formed by the serial concatenation of a repetition code with multiple accumulators, and the hybrid concatenated code (HCC) ensembles recently introduced by Koller *et al.* (*5th Int. Symp. on Turbo Codes & Rel. Topics*, Lausanne, Switzerland) consisting of an outer multiple parallel concatenated code serially concatenated with an inner accumulator. We introduce stopping sets for iterative constituent code oriented decoding using maximum *a posteriori* erasure correction in the constituent codes. We then analyze the asymptotic stopping set distribution for RMA and HCC ensembles and show that their stopping distance $h_{\min}$, defined as the size of the smallest nonempty stopping set, asymptotically grows linearly with the block length. Thus, these code ensembles are *good* for the BEC. It is shown that for RMA code ensembles, contrary to the asymptotic minimum distance $d_{\min}$, whose growth rate coefficient increases with the number of accumulate codes, the $h_{\min}$ growth rate coefficient diminishes with the number of accumulators. We also consider random puncturing of RMA code ensembles and show that for sufficiently high code rates, the asymptotic $h_{\min}$ does not grow linearly with the block length, contrary to the asymptotic $d_{\min}$, whose growth rate coefficient approaches the Gilbert-Varshamov bound as the rate increases. Finally, we give iterative decoding thresholds for the different code ensembles to compare the convergence properties.

*Index Terms*—Asymptotic stopping set distribution, binary erasure channel, EXIT charts, hybrid concatenated codes, repeat accumulate codes, spectral shape function, stopping set, uniform interleaver.

## I. INTRODUCTION

Turbo codes introduced by Berrou *et al.* in [1], and Gallager's low-density parity-check (LDPC) codes [2], rediscovered by MacKay in [3], are considered among the most powerful error-correction schemes of today due to their low decoding complexity and near-capacity performance on a wide variety of channels. Recently, serially concatenated codes, such as very simple repeat accumulate (RA) codes, and hybrid concatenated codes (HCCs), i.e., mixed parallel and serial structures combining the features of the two concatenations [4–13], have also attracted some attention, since they yield better minimum distances than turbo codes.

For repeat multiple-accumulate (RMA) codes, Pfister and Siegel [5] showed that the minimum distance $d_{\min}$ increases

Manuscript received October 1, 2008 and revised January 15, 2009.
A. Graell i Amat is with the Department of Electronics, Institut TELECOM-TELECOM Bretagne, CS 83818 - 29238 Brest Cedex 3, France (e-mail: alexandre.graell@telecom-bretagne.eu). E. Rosnes is with the Selmer Center, Department of Informatics, University of Bergen, N-5020 Bergen, Norway (e-mail: eirik@ii.uib.no). A. Graell i Amat is supported by a Marie Curie Intra-European Fellowship within the 6th European Community Framework Programme. The work of E. Rosnes was supported by the Norwegian Research Council (NFR) under Grants 174982 and 183316.

as the number of accumulators increase. In the limit of infinitely many accumulators, it was shown in [5] that the $d_{\min}$ approaches the Gilbert-Varshamov bound (GVB). Recently, Kliewer *et al.* [10] and Fagnani and Ravazzi [11, 12] showed independently that RMA code ensembles are asymptotically good, in the sense that their typical $d_{\min}$ asymptotically grows linearly with the block length. A method for the calculation of a lower bound on the growth rate coefficient was also given in [10]. The same principle was applied in [8, 9] to HCC structures. In this work, when we speak about HCCs we mean the HCC structures from [8], even if this is not explicitly stated.

In iterative decoding for the binary erasure channel (BEC), stopping set distributions play an analogous role to that of the distance spectra in maximum-likelihood (ML) decoding. Stopping sets for iterative belief-propagation (BP) decoding of LDPC codes were introduced by Di *et al.* in [14]. In [15], Rosnes and Ytrehus adapted the concept of stopping sets to turbo decoding and introduced turbo stopping sets. Also, an exact condition for decoding failure on the BEC was stated as follows. Apply turbo decoding until the transmitted codeword has been recovered, or the decoder fails to progress further. Then the set of erased positions that remain when the decoder stops is equal to the unique maximum-size turbo stopping set which is also a subset of the (initial) set of erased positions. The stopping set concept has also been adapted to iterative row-column decoding of product codes in a recent paper [16].

In this work, we adapt the concept of stopping sets to iterative constituent code oriented decoding using maximum *a posteriori* (MAP) erasure correction in the constituent codes of RMA codes and HCCs. We then give expressions for their average stopping set distributions and analyze their asymptotic behavior. We show that both RMA and HCC ensembles, consisting of an outer multiple parallel concatenated code (MPCC) serially concatenated with an inner accumulator [8], exhibit an asymptotic stopping distance $h_{\min}$, defined as the size of the smallest nonempty stopping set [17], that grows linearly with the block length. Therefore, these code ensembles are *good* for the BEC. For RMA code ensembles, contrary to the asymptotic $d_{\min}$, whose growth rate coefficient increases with the number of accumulate codes, the asymptotic $h_{\min}$ growth rate coefficient diminishes with the number of accumulators. We also consider random puncturing of the RMA code ensemble and show that for sufficiently high code rates, the asymptotic $h_{\min}$ does not grow linearly with the block length, contrary to the asymptotic $d_{\min}$, whose growth rate coefficient approaches the GVB as the rate increases [10].

The reminder of this paper is organized as follows. The encoder structures of RMA codes and the HCCs from [8] are described in Section II. In Section III, we describe iterative constituent code oriented decoding using MAP erasure correction in the constituent codes of concatenated codes and introduce stopping sets for this particular iterative decoding algorithm. We further give some of the basic properties and show that these stopping sets characterize exactly the performance of iterative constituent code oriented decoding on the BEC. Section IV discusses stopping set enumerators for RMA and HCC ensembles. Also, a factor graph interpretation is given. Section V introduces the asymptotic stopping set size spectral shape function. In Sections VI and VII a stopping distance analysis is performed for RMA code ensembles and HCC ensembles, respectively. Convergence properties are studied in Section VIII, where an extrinsic information transfer (EXIT) charts analysis is performed. Conclusions are given in Section IX.

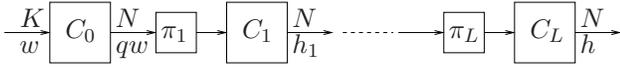

Fig. 1. Encoder structure for RMA codes.

## II. ENCODER STRUCTURES

The encoder structure of RMA codes is depicted in Fig. 1. It is the serial concatenation of a repetition code $C_0$ of rate $R_0 = 1/q$ with the cascade of $L \geq 1$ identical rate-1, memory-one, accumulators $C_l$, $l = 1, \ldots, L$, with generator polynomials $g(D) = 1/(1 + D)$, through interleavers $\pi_1, \ldots, \pi_L$. The overall nominal code rate (avoiding termination) is denoted by $R = K/N = 1/q$, where $K$ and $N$ denote the input and the output block length, respectively. Higher rates may be obtained by puncturing the output of the most inner accumulator $C_L$.

We also consider the HCC ensembles introduced in [8] consisting of an outer MPCC with $q$ parallel branches $C_l$, $l = 1, \ldots, q$, serially concatenated with an inner accumulator, denoted by $C_{q+1}$. Four different encoder structures, depicted in Fig. 2, are considered. For type-1 and type-2 codes, all the code bits from the outer MPCC enter the inner accumulator, while for type-3 and type-4 codes only $q - 1$ of the $q$ parallel branches enter the inner accumulator. The nominal code rate is denoted by $R = K/N$. For type-1, type-3, and type-4 HCCs the code rate is $R = 1/q$, while for type-2 HCCs $R = 1/(q_1 + q_2) = 1/q$, where $q_1$ denotes the number of feedforward branches and $q_2$ denotes the number of recursive branches. The outer MPCC of type-2 and type-3 HCCs generalizes the $R = 1/4$ MPCC introduced in [18], which incorporates a feedforward branch ($g(D) = 1+D$) since it yields better convergence behavior than a MPCC with only recursive branches. Note also that in the type-4 HCC encoder $C_1$ performs an identity mapping, i.e., the HCC is systematic. In [8], the asymptotic and finite-length $d_{\min}$ properties of these four encoders were studied in detail for $q = 4$. Furthermore, iterative decoding thresholds on the additive white Gaussian noise (AWGN) channel were estimated using EXIT charts. In terms of $d_{\min}$ properties, type-1 codes are the best and type-4 codes the worst, while in terms of iterative decoding thresholds on the AWGN channel the ranking is opposite [8].

## III. ITERATIVE CONSTITUENT CODE ORIENTED DECODING AND STOPPING SETS

In this paper, we consider stopping sets for iterative decoding of concatenated codes. Thus, we assume that the concatenated codes in Figs. 1 and 2 are decoded iteratively in a constituent code oriented fashion using MAP constituent decoders over the BEC. By iterative constituent code oriented decoding we mean a decoding strategy that iterates between the constituent decoders, i.e., the turbo-decoding principle applied to more general concatenated codes. MAP decoding of the constituent codes can be implemented efficiently on a trellis representation of the constituent code [15, 19]. The aim of trellis-based iterative constituent code oriented decoding on the BEC is to find a set of paths through each constituent code trellis that is consistent with the received sequence. The decoding starts with a set of all paths and iteratively eliminates those that are inconsistent. This iterative process continues until either there is only one possible path left in each constituent trellis (successful decoding), or there is no change from one iteration step to the next [15]. The complexity of iterative constituent code oriented decoding of concatenated codes is linear in the block length $N$ when the number of constituent code activations is independent of $N$.

As an alternative to trellis-based decoding, decoding of the constituent codes could be based, e.g., on a factor graph representation, but this will not necessarily be MAP erasure correction. Note that if the factor graph is constructed based on all codewords of the dual code, or if the factor graph does not contain any cycles, then the *peeling decoder* of Luby *et al.* [20] implements MAP erasure correction in the constituent codes [21, 22]. Indeed, for the stopping set analysis carried out in this paper, we do not require to make any assumption on the decoding algorithm used at each constituent decoder. The key point is that decoding is performed in a constituent code oriented fashion, and that the constituent decoders perform MAP decoding. However, we prefer to refer to trellis-based decoding, since the trellis representation of the constituent encoders is very useful for the derivation of the stopping set enumerators and the subsequent asymptotic analysis. This will become apparent in the following sections.

### A. Stopping Sets for RMA Codes

We will now give the formal definition of a stopping set for RMA codes, adapted from the definition in [15] for turbo stopping sets. The generalization to the HCC ensembles of Fig. 2 and to the case with puncturing is straightforward. In the following stopping set definition, we need the concept of support set of a binary vector and of a binary linear code. The support set $\chi(\mathbf{x})$ of a binary vector $\mathbf{x} = (x_1, \ldots, x_N)$ (of length $N$) is the set of nonzero coordinates. As an example, with $\mathbf{x} = (0, 1, 1, 0, 1)$, $\chi(\mathbf{x}) = \{2, 3, 5\}$. The support set $\chi(C)$ of a binary linear code $C$ is the union of the support sets of each codeword in $C$. Also, we need the concept of a *subcode*. A subcode $\tilde{C}$ of a binary linear code $C$ is a subspace of $C$. Finally, an interleaver will be regarded as a mapping from the set of coordinates of its input sequence to the set of coordinates of its output sequence.

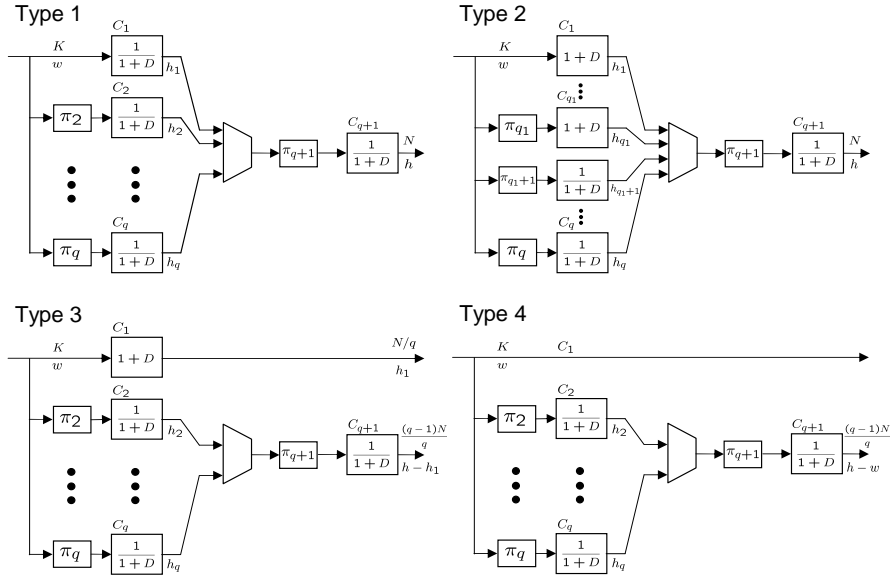

Fig. 2. Encoder structures for HCCs.

*Definition 1:* Let $C_{\text{RMA}}$ denote a given RMA code with interleavers $\pi_1, \ldots, \pi_L$. A set $\mathcal{S} = \mathcal{S}(\pi_1, \ldots, \pi_L) \subseteq \{1, \ldots, N\}$ of the coordinates of the output sequence (or codeword) is a stopping set if and only if there exist $L+1$ linear subcodes $\hat{C}_l \subseteq C_l \subseteq \{0,1\}^N$, $l = 0, \ldots, L$, with support sets $\chi(\hat{C}_l)$ such that

$$\chi(\hat{C}_L) = \mathcal{S} \text{ and } \pi_l(\chi(\hat{C}_{l-1})) = \chi(\hat{C}_l),\ l = 1, \ldots, L.$$

The *size* of a stopping set $\mathcal{S}$ is its cardinality.

Note that in Definition 1 we used the fact that the mapping between the input support set and the output support set is an identity mapping for rate $R = 1$ encoders. For general $R < 1$ encoders this is not true. Also, note that Definition 1 does not exclude the empty set. Thus, the empty set is formally a stopping set of size zero. The size of the smallest nonempty stopping set is called the *stopping distance* [17] and is denoted by $h_{\min}$. We emphasize the fact that the concept of stopping sets for RMA codes, as defined above, is conceptually different from the traditional concept of stopping sets used in connection with iterative BP decoding of LDPC codes, but it reduces to the traditional concept of stopping sets for Tanner graphs when the constituent codes are single parity-check codes. Indeed, the RMA code could be decoded using iterative BP decoding on a Tanner graph representing the entire code [23]. On the other hand, the concept of stopping sets from Definition 1 is related to iterative constituent code oriented decoding. In general, the concept of stopping sets can be defined for *any* iterative decoding algorithm when it operates on the BEC and should not be used without having a specific iterative decoding algorithm in mind.

In general, the erasure probability after iterative BP decoding of a concatenated code using its factor graph is greater than or equal to the erasure probability after iterative constituent code oriented decoding using MAP erasure correction in the constituent codes for any channel erasure pattern. Here, with the factor graph of a general concatenated code we mean a factor graph constructed from arbitrary Tanner graphs of each constituent code. The Tanner graphs of the constituent codes are interconnected through the interleavers. An example of such a factor graph is given in Fig. 4 for a repeat accumulate-accumulate (RAA) code with $q = 3$. The interested reader is referred to [15] for further details. For the special case of RMA codes and HCCs with accumulate constituent encoders, it turns out that iterative BP decoding on the overall factor graph is equivalent to iterative constituent code oriented decoding. This is because the factor graph of an accumulator (depicted in Fig. 4) does not contain cycles [24, p. 583]. Thus, iterative BP decoding of a constituent accumulator (using a factor graph) is equivalent to MAP erasure correction. Actually, this result holds for any accumulator with generator polynomial $1/(1 + D^t)$, $t \geq 1$. However, we stress that the traditional definition of stopping sets (on an overall factor graph) is only appropriate when the constituent encoders are accumulators, but not for general concatenated codes, i.e., when other convolutional codes are used as constituent codes. This is one of the reasons why we have introduced iterative constituent code oriented decoding and the corresponding general definition of stopping sets in Definition 1. Indeed, an asymptotic stopping distance analysis of the concatenated codes analyzed here using, e.g., 4-state or 8-state convolutional encoders could be carried out using the approach in [25] to compute asymptotic input-output weight distributions of convolutional encoders, generalized to the stopping set case.

In summary, the stopping set enumerators and the analysis in the following sections are general and do not only apply to iterative constituent code oriented decoding; iterative BP decoding on the overall factor graph of RMA codes and the HCCs of Fig. 2 leads to the same stopping set enumerators as iterative constituent code oriented decoding.

The following lemma can easily be proved.

*Lemma 1:* Let $C$ denote a RMA code (or a HCC). Then, the support set of any codeword in $C$ is a stopping set.

*Theorem 1:* Let $C$ denote a RMA code (or a HCC) that we use to transmit information over the BEC. The received vectors are decoded iteratively using constituent code oriented decoding using MAP erasure correction in the constituent codes until either the codeword has been recovered, or the decoder fails to progress further. Then the set of erased positions that remain when the decoder stops is equal to the unique maximum-size stopping set that is contained in the (initial) set of erased positions.

*Proof:* The proof uses the same basic ideas as the proof of Theorem 1 in [15], and is omitted for brevity. ∎

From Theorem 1 it arises that an important parameter for code performance is the stopping distance $h_{\min}$. In the following sections we address the asymptotic behavior of $h_{\min}$ for RMA and HCC ensembles.

## IV. Support Sets and Stopping Set Enumerators

Let $C$ denote an $(N, K)$ binary linear code. Partition all the subcodes of $C$ of dimension $d$, $d = 0, \ldots, K$, into equivalence classes based on their support sets. In particular, all subcodes within a specific subcode class are required to have the same support set, but the subcodes may have different dimensions. We define the *subcode input-output support size enumerating function* (SIOSEF) [15] of $C$ as

$$A^C(W, H) = \sum_{w=0}^{K} \sum_{h=0}^{N} a_{w,h}^C W^w H^h$$

where $W$ and $H$ are dummy variables, and $a_{w,h}^C$ is the number of subcode classes of $C$ of *input* support set size $w$ and *output* support set size $h$. In the rest of the paper, with a slight abuse of notation, we will refer interchangeably to both $A^C(W, H)$ and $a_{w,h}^C$ as the SIOSEF of a code $C$.

### A. Stopping Set Enumerators for RMA and HCC Ensembles

Let $\bar{s}_{w,h}^{\mathcal{C}}$ be the ensemble-average *input-output stopping set size enumerating function* (IOSSEF) of the code ensemble $\mathcal{C}$ with input and output block length $K$ and $N$, respectively, denoting the average number of stopping sets of input size $w$ and output size $h$ over $\mathcal{C}$. Also, denote by $\bar{s}_h^{\mathcal{C}} = \sum_{w=0}^{K} \bar{s}_{w,h}^{\mathcal{C}}$ the ensemble-average *stopping set size enumerating function* (SSEF) of the code ensemble $\mathcal{C}$, giving the average number of stopping sets of size $h$ over $\mathcal{C}$.

Using the concept of uniform interleaver [26], the IOSSEF of a RMA code ensemble $\mathcal{C}_{\text{RMA}}$ can be obtained from the SISOEFs of the constituent encoders. Indeed, the contribution of the SISOEFs of the constituent encoders to the IOSSEF of a RMA code ensemble is, through Definition 1, analogous to the role played by the weight spectra of constituent encoders in the weight spectrum of a concatenated code ensemble. The ensemble-average IOSSEF of $\mathcal{C}_{\text{RMA}}$ can then be written as [15, 26]

$$\bar{s}_{w,h}^{\mathcal{C}_{\text{RMA}}} = \sum_{h_1=0}^{N} \cdots \sum_{h_{L-1}=0}^{N} \frac{a_{w,qw}^{C_0} a_{qw,h_1}^{C_1}}{\binom{N}{qw}} \left[ \prod_{l=2}^{L-1} \frac{a_{h_{l-1},h_l}^{C_l}}{\binom{N}{h_{l-1}}} \right] \frac{a_{h_{L-1},h}^{C_L}}{\binom{N}{h_{L-1}}}$$

$$= \sum_{h_1=0}^{N} \cdots \sum_{h_{L-1}=0}^{N} \bar{s}_{w,h_1,\ldots,h_{L-1},h}^{\mathcal{C}_{\text{RMA}}}$$

(1)

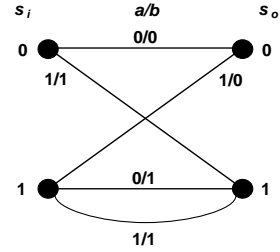

Fig. 3. Extended trellis module of the rate-1 recursive convolutional encoder with generator polynomial $g(D) = 1/(1 + D)$. The edge labels have the format input/output.

where $\bar{s}_{w,h_1,\ldots,h_{L-1},h}^{\mathcal{C}_{\text{RMA}}}$ is called the conditional support size enumerating function of $\mathcal{C}_{\text{RMA}}$.

Now, consider the $(N, K)$ HCC ensembles $\mathcal{C}_{\text{HCC}}$ of Fig. 2. Assuming trellis termination, the $l$th constituent code is a binary linear code with input block length $K_l$ and output block length $N_l$. Except for $C_1$, which is directly connected to the input, every code is preceded by an interleaver. Furthermore, we assume that the code $C_{q+1}$ is directly connected to the channel. Partition the set $\{1, 2, \ldots, q\}$ into two disjoint subsets $Q$ and its complement $\bar{Q}$ such that $Q$ contains the indices of all codes directly connected to the channel. The ensemble-average IOSSEF of $\mathcal{C}_{\text{HCC}}$ can be written as [8, 15]

$$\bar{s}_{w,h}^{\mathcal{C}_{\text{HCC}}} = \sum_{h_1=0}^{N_1} \cdots \sum_{h_q=0}^{N_q} \frac{a_{w,h_1}^{C_1} a_{\sum_{l \in \bar{Q}} h_l, h - \sum_{l \in Q} h_l}^{C_{q+1}}}{\binom{K_{q+1}}{\sum_{l \in \bar{Q}} h_l}} \prod_{l=2}^{q} \frac{a_{w,h_l}^{C_l}}{\binom{K_l}{w}}$$

$$= \sum_{h_1=0}^{N_1} \cdots \sum_{h_q=0}^{N_q} \bar{s}_{w,h_1,\ldots,h_q,h}^{\mathcal{C}_{\text{HCC}}}$$

(2)

where $\bar{s}_{w,h_1,\ldots,h_q,h}^{\mathcal{C}_{\text{HCC}}}$ is called the conditional support size enumerating function of $\mathcal{C}_{\text{HCC}}$.

The evaluation of (1) and (2) requires the computation of the SIOSEFs of the constituent encoders. In the following, we give closed-form expressions for the SIOSEFs of rate-1, memory-one, encoders, and repeat codes. They will be used later to derive the asymptotic expressions for the stopping set distributions in Sections VI and VII.

### B. SIOSEFs for Memory-One Encoders and the Repetition Code

*Theorem 2:* The SIOSEF for rate-1, memory-one, convolutional encoders with generator polynomials $g(D) = 1/(1+D)$ and $g(D) = 1 + D$ that are terminated to the zero state at the end of the trellis and with input and output block length $N$ can be given in closed form as

$$a_{w,h}^{\frac{1}{1+D}} = a_{h,w}^{1+D} = \sum_{d=1}^{\lfloor \frac{w}{2} \rfloor} \binom{N-h}{d} \binom{h-1}{d-1} \binom{h-d}{w-2d} \quad (3)$$

for positive input sizes $w$. Also, $a_{0,0}^{\frac{1}{1+D}} = a_{0,0}^{1+D} = 1$.

*Proof:* Let $C^N$ denote the binary linear code obtained by terminating (to the zero state) the rate-1 recursive convolutional encoder with generator polynomial $g(D) = 1/(1 + D)$ and with input and output block length of $N$. In Fig. 3, the *extended* trellis module [15], denoted by $\mathcal{T}_\text{ext}$, of the encoder is depicted. We call the trellis module in Fig. 3 the extended trellis module because it extends (and comprises) the trellis module of the code to represent all support vectors of subcodes. Each edge between two trellis states is labeled with a binary pair $a/b$, where $a$ denotes the input support vector bit and $b$ denotes the output support vector bit. The overall transition between two states $s_i$ and $s_o$ is denoted by $(s_i, a/b, s_o)$. Note that $\mathcal{T}_\text{ext}$ is identical to the trellis module of the original code, except that it has an extra edge (with label $1/1$) from state one to state one. The paths of length $N$ starting from the zero state and ending in the zero state in an $N$-fold concatenation of $\mathcal{T}_\text{ext}$ give all possible support vectors of subcodes of $C^N$ [15]. Also, the $N$-fold concatenation of $\mathcal{T}_\text{ext}$ does not contain duplicates, i.e., there are no two paths with the same input/output label sequence. From [27], we know that the number of paths in an $N$-fold concatenation of $\mathcal{T}_\text{ext}$ corresponding to codewords, i.e., paths that do not contain the $(1, 1/1, 1)$ transition, of even input weight $w \geq 2$ (there are no codewords of odd input weight) and output weight $h$ is

$$\binom{N-h}{w/2}\binom{h-1}{w/2-1}.$$

These codewords consist of $w/2$ error events with a total of $h - w/2$ transitions from state one to state one. Now, the number of paths in the extended trellis, i.e., the $N$-fold concatenation of $\mathcal{T}_\text{ext}$, consisting of $d$ error events with a total output weight of $h$ and total input weight of $w \geq 2$ is

$$\binom{N-h}{d}\binom{h-1}{d-1}\binom{h-d}{w-2d}$$

since the transition $(1, 1/1, 1)$ does not increase the output weight ($b$ is 1 for both edges between state one and state one), but only the input weight. The result for the encoder $g(D) = 1/(1+D)$ follows by summing over all possible values of $d$. Furthermore, $a_{0,0}^{\frac{1}{1+D}} = 1$, since the empty set is the support set of the trivial subcode containing only the all-zero codeword. The SIOSEF for the feedforward encoder with generator polynomial $g(D) = 1 + D$ is obtained in a similar manner. ∎

Note that the formula in (3) generalizes the closed-form expression for the input-output weight enumerating function for rate-1, memory-one, convolutional encoders from [27].

*Theorem 3:* The SIOSEF for the $(Kq, K)$ repetition code $C_0$ with input block length $K$ obtained by concatenating together $K$ successive codewords of a $(q, 1)$ repetition code can be given in closed form as

$$a_{w,qw}^{C_0} = \binom{K}{w}. \qquad (4)$$

*Proof:* Consider the $(q, 1)$ repetition code $C$. This code consists of two codewords, the all-zero codeword and the all-one codeword. Obviously, $A^C(W, H) = 1 + WH^q$. With an input block length of $K$, we get $(1 + WH^q)^K = \sum_{w=0}^{K}\binom{K}{w}W^w H^{qw}$, from which it follows that $a_{w,qw}^{C_0} = \binom{K}{w}$. ∎

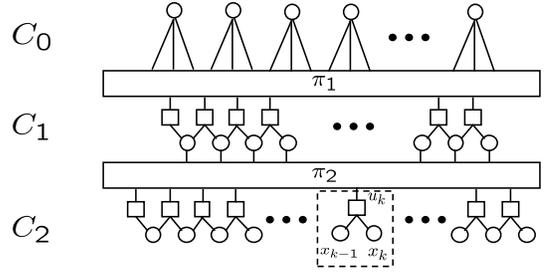

Fig. 4. Factor graph representation of a RAA code with $q = 3$.

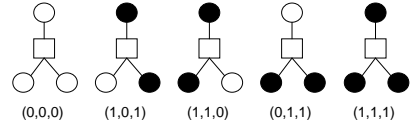

(0,0,0)   (1,0,1)   (1,1,0)   (0,1,1)   (1,1,1)

Fig. 5. Factor graph representation of an extended trellis section of an accumulate code.

### C. A Factor Graph Interpretation

In this section, we consider an equivalent factor graph representation of the accumulate code to better clarify the computation of the SIOSEF and its connection with stopping sets of a concatenated code. The factor graph of a RAA code with $q = 3$ is depicted in Fig. 4. Consider a single section of the factor graph corresponding to one of the two accumulate codes, marked with a dashed box, and transmission over the BEC. We have the relation $u_k = x_{k-1} + x_k$, where $u_k$ is the information symbol, and $x_{k-1}$ and $x_k$ are the code symbols of section $k$. The decoder will fail in recovering an erased symbol at section $k$ if and only if two or more symbols at section $k$ are erased. Four possible triplets $(u_k, x_{k-1}, x_k)$ are possible, namely $(1, 0, 1)$, $(1, 1, 0)$, $(0, 1, 1)$, and $(1, 1, 1)$, where 1 means that the symbol is erased. Their factor graph representation is given in Fig. 5, where a black circle means that the symbol is erased (1) and an empty circle means that the symbol is not erased (0). Clearly, to enumerate the *stopping sets* of an accumulate code of length $N$ we must consider all valid combinations (of length $N$) of the sections in Fig. 5 plus the one corresponding to the triplet $(0, 0, 0)$ (no erasures). Note that we do not need to consider the triplets with a single one, since in this case the erasure would be recovered, yielding to $(0, 0, 0)$. To simplify this computation, the five factor graph sections of Fig. 5 can be equivalently represented in a more compact form by $\mathcal{T}_\text{ext}$ in Fig. 3, which gives a trellis representation of all possible support vectors of subcodes of an accumulate code $C$. The five triplets $(0, 0, 0)$, $(1, 0, 1)$, $(1, 1, 0)$, $(0, 1, 1)$, and $(1, 1, 1)$ of Fig. 5 correspond to transitions $(0, 0/0, 0)$, $(0, 1/1, 1)$, $(1, 1/0, 0)$, $(1, 0/1, 1)$, and $(1, 1/1, 1)$, respectively, in $\mathcal{T}_\text{ext}$. The enumeration of all *stopping sets*, or all support sets of subcodes, of the accumulate code, given by $a_{w,h}^{\frac{1}{1+D}}$, can then be computed using $\mathcal{T}_\text{ext}$ as shown in the previous subsection. Please note that here, for better understanding, we used the concept of *stopping sets* applied to an accumulate code. Formally speaking, stopping sets, as defined in Section III-A, apply only to iterative decoding of concatenated codes. When referring to the constituent codes (with the associated MAP decoding algorithm) it is more appropriate to use the concept of support sets discussed in

Section III-A.

Finally, the IOSSEF of RMA and HCC ensembles can be obtained by properly combining the SIOSEFs of the constituent encoders through (1) and (2), respectively.

## V. FINITE-LENGTH AND ASYMPTOTIC ANALYSIS OF STOPPING SETS

### A. Finite-Length Analysis of Stopping Sets

The ensemble-average SSEF $\bar{s}_h^{\mathcal{C}}$ of a concatenated code ensemble $\mathcal{C}$ can be used to bound its $h_{\min}$ in the finite-length regime, similarly to the case of the $d_{\min}$ [2, 5]. In particular, the following bound holds.

*Lemma 2:* The probability that a code chosen randomly from the ensemble $\mathcal{C}$ with ensemble-average SSEF $\bar{s}_h^{\mathcal{C}}$ has stopping distance $h_{\min} < \hbar$ is upper-bounded as

$$\Pr(h_{\min} < \hbar) \leq \sum_{h=1}^{\hbar-1} \bar{s}_h^{\mathcal{C}}. \quad (5)$$

*Proof:* The bound follows from the application of the union bound and the Markov inequality [5]. ∎

Lemma 2 can be used to obtain a probabilistic lower bound on the stopping distance of a code ensemble. In particular, if we set $\Pr(h_{\min} < \hbar) = \varepsilon$, where $\varepsilon$ is any positive value between 0 and 1, we would expect that at least a fraction $1 - \varepsilon$ of the codes in the ensemble have a stopping distance $h_{\min}$ of at least $\hbar$.

### B. Asymptotic Analysis of Stopping Sets

We define the asymptotic stopping set size spectral shape function as [2]

$$r_{\mathrm{s}}(\rho) = \lim_{N \longrightarrow \infty} \sup \frac{1}{N} \ln \bar{s}_{\lfloor \rho N \rfloor}^{\mathcal{C}} \quad (6)$$

where $\sup(\cdot)$ denotes the supremum of its argument, $\rho = \frac{h}{N}$ is the normalized stopping set size, and $N$ is the block length. From (6), the SSEF can be expressed as $\bar{s}_h^{\mathcal{C}} \sim e^{N r_{\mathrm{s}}(\rho)}$ when $N \longrightarrow \infty$. Therefore, if there exists some abscissa $\rho_0 > 0$ such that $\sup_{\rho \leq \rho^*} r_{\mathrm{s}}(\rho) < 0 \quad \forall \rho^* < \rho_0$, and $r_{\mathrm{s}}(\rho) > 0$ for some $\rho > \rho_0$, then it can be shown (using Lemma 2 for example) that, with high probability, the stopping distance of most codes in the ensemble grows linearly with the block length $N$, with growth rate coefficient of at least $\rho_0$. On the other hand, if $r_{\mathrm{s}}(\rho)$ is strictly zero in the range $(0, \rho_0)$, it cannot be proved directly whether $h_{\min}$ grows linearly with the block length or not.

In [11, 12], it was shown that the spectral shape function of RMA codes, for the codeword case, exhibits this behavior, i.e., it is zero in the range $(0, \rho_0)$ and positive for some $\rho > \rho_0$, where $\rho$ means here the normalized output weight. By combining the asymptotic spectral shapes with the use of bounding techniques, Fagnani and Ravazzi were able to prove in [11, Theorem 11] that the minimum distance of RMA codes indeed grows linearly with the block length with growth rate coefficient of at least $\rho_0$. In Sections VI and VII we show that the stopping set size spectral shape function of RMA and HCC ensembles exhibits a similar behavior. We shall then extend Theorem 11 in [11] to prove that RMA codes and HCCs are also good for the BEC.

We remark that in the rest of the paper, with a slight abuse of language, we sometimes refer to $\rho_0$ as the exact value of the asymptotic growth rate coefficient. However, we emphasize that, strictly speaking, $\rho_0$ is only a lower bound on the asymptotic growth rate coefficient.

## VI. STOPPING DISTANCE ANALYSIS FOR RMA CODES

In this section, we analyze the asymptotic stopping distance properties of RMA code ensembles. We first show that the $h_{\min}$ of RA code ensembles cannot grow linearly with the block length. Then, we consider the ensemble of RMA codes obtained from the concatenation of a repeat code with the cascade of $L > 1$ accumulators and show that the asymptotic $h_{\min}$ grows linearly with the block length.

### A. RA Codes

We consider the $h_{\min}$ of code ensembles formed by the concatenation of a $R_0 = 1/q$ repetition code with a single inner accumulator ($L = 1$) with generator polynomial $g(D) = 1/(1+D)$ through a uniform interleaver [26] (see Fig. 1). In particular, we prove the following theorem.

*Theorem 4:* The stopping distance $h_{\min}^{\mathcal{C}_{\mathrm{RA}}}$ of a RA code ensemble $\mathcal{C}_{\mathrm{RA}}$ with repetition factor $q \geq 3$ satisfies

$$\lim_{N \longrightarrow \infty} \Pr\left(h_{\min}^{\mathcal{C}_{\mathrm{RA}}} \leq N^{\frac{q-2}{3q+2} - \epsilon}\right) = 0$$

where $\epsilon$ is any positive constant.

*Proof:* Using (1), (3), and (4), the ensemble-average IOSSEF $\bar{s}_{w,h}^{\mathcal{C}_{\mathrm{RA}}}$ with $w > 0$ of a RA code ensemble can be written as

$$\bar{s}_{w,h}^{\mathcal{C}_{\mathrm{RA}}} = \frac{\binom{K}{w}\sum_{d=1}^{\lfloor \frac{qw}{2} \rfloor}\binom{N-h}{d}\binom{h-1}{d-1}\binom{h-d}{qw-2d}}{\binom{N}{qw}}.$$

Using Stirling's approximation $\binom{n}{k} \leq \left(\frac{ne}{k}\right)^k$ and the fact that $\prod_{i=0}^{l}(N-i) \geq \frac{N^{l+1}}{\varphi_N(l)}$, with $\varphi_\lambda(l) = \exp\left(\frac{l(l+1)}{2\lambda}\right)$ [10], $\bar{s}_{w,h}^{\mathcal{C}_{\mathrm{RA}}}$ can be bounded as

$$\bar{s}_{w,h}^{\mathcal{C}_{\mathrm{RA}}} \leq N^{w - \lceil \frac{qw}{2} \rceil} h^{qw + \lfloor \frac{qw}{2} \rfloor - 3}$$

$$\times \underbrace{\frac{(qw)! e^{qw + w - 1} \varphi_N(qw - 1)}{q^w w^w} \sum_{d=1}^{\lfloor \frac{qw}{2} \rfloor} \frac{(qw - 2d)^{2d - qw}}{d^d (d-1)^{d-1}}}_{g(w, N)}.$$

(7)

The total average number of nonempty stopping sets of size $h \leq \hbar$ can now be obtained from $\bar{s}_{w,h \leq \hbar}^{\mathcal{C}_{\mathrm{RA}}} = \sum_{h=1}^{\hbar} \bar{s}_{w,h}^{\mathcal{C}_{\mathrm{RA}}}$. Using (7), we get

$$\bar{s}_{w,h \leq \hbar}^{\mathcal{C}_{\mathrm{RA}}} \leq N^{w - \lceil \frac{qw}{2} \rceil} \hbar^{qw + \lfloor \frac{qw}{2} \rfloor} g(w, N). \quad (8)$$

Let $\bar{s}_{h \leq \hbar}^{\mathcal{C}_{\mathrm{RA}}} = \sum_{w=1}^{K} \bar{s}_{w,h \leq \hbar}^{\mathcal{C}_{\mathrm{RA}}} = \sum_{w=1}^{2\hbar/q} \bar{s}_{w,h \leq \hbar}^{\mathcal{C}_{\mathrm{RA}}}$, where we used the fact that $w \leq 2\hbar/q$. This follows from the binomial coefficient $\binom{h-d}{qw-2d}$, from which it follows that $qw \leq h + d \leq h + \lfloor qw/2 \rfloor$, which implies that $w \leq 2h/q \leq 2\hbar/q$. Also, we assume that $\hbar$ can be expressed in terms of the block length $N$ as $\hbar = N^\nu$, with $0 < \nu < 1$. From (8), it follows that

$$\bar{s}_{h \leq \hbar}^{\mathcal{C}_{\text{RA}}} \leq \frac{2}{q} \max_{1 \leq w \leq 2h/q} N^{w - \lceil \frac{qw}{2} \rceil} \hbar^{qw + \lfloor \frac{qw}{2} \rfloor + 1} g(w, N)$$

$$= \frac{2}{q} N^{1 - \lceil \frac{q}{2} \rceil} \hbar^{q + \lfloor \frac{q}{2} \rfloor + 1} g(1, N)$$

for large enough $N$. We choose $\nu$ such that

$$\lim_{N \to \infty} N^{1 - \lceil \frac{q}{2} \rceil} \hbar^{q + \lfloor \frac{q}{2} \rfloor + 1} g(1, N) = 0$$

which implies that

$$1 - \left\lceil \frac{q}{2} \right\rceil + \nu \left( q + \left\lfloor \frac{q}{2} \right\rfloor + 1 \right) < 0 \tag{9}$$

since $g(1, N)$ approaches a constant when $N$ approaches infinity. The evaluation of (9) leads to $\nu < \frac{q-2}{3q+2}$, and we can express the stopping set size $\hbar$ as $N^{\frac{q-2}{3q+2} - \epsilon}$ for any positive constant $\epsilon$. Since (see Lemma 2 in Section V-A)

$$\Pr\left( h_{\min}^{\mathcal{C}_{\text{RA}}} \leq \hbar \right) \leq \bar{s}_{h \leq \hbar}^{\mathcal{C}_{\text{RA}}}$$

the theorem follows. ∎

We remark that we can improve the lower bound of $(q-2)/(3q+2)$ from Theorem 4 using a much more convoluted approach. In particular, the lower bound can be tightened to $(q-2)/(3q)$ for all $q \geq 3$ using a similar technique as in the proof of Theorem 1 in [10]. However, in the stopping set case, the summation over $d$ in (7) needs to be upper-bounded by $\lfloor \frac{qw}{2} \rfloor$ times the maximum element found by taking the derivative with respect to $d$. Further details are omitted for brevity.

*Corollary 1:* In the ensemble of RA codes with repetition factor $q \geq 3$ almost all codes have stopping distance lower-bounded for $N \longrightarrow \infty$ by

$$h_{\min}^{\mathcal{C}_{\text{RA}}} > N^{\frac{q-2}{3q+2} - \epsilon}$$

where $\epsilon$ is any positive constant.

An upper bound for the stopping distance is given by the following theorem.

*Theorem 5:* The stopping distance $h_{\min}^{\mathcal{C}_{\text{RA}}}$ of *any* code in the RA code ensemble with repetition factor $q \geq 3$ is upper-bounded for $N \longrightarrow \infty$ by

$$h_{\min}^{\mathcal{C}_{\text{RA}}} \leq O\left( N^{\frac{q-1}{q}} \right).$$

*Proof:* The proof given in [28] for $d_{\min}$ of RA code ensembles still holds for the stopping distance. ∎

As a consequence of Theorems 4 and 5, the stopping distance of almost all codes in the RA code ensemble with $q \geq 3$ grows with $N$ as $O(N^\nu)$, where $(q-2)/(3q+2) \leq \nu \leq (q-1)/q$, i.e., the RA code ensemble is *bad* for the BEC.

### B. RMA Codes

In the following, we consider the ensemble of codes formed by a rate $R_0 = 1/q$ repetition code followed by the cascade of $L$ accumulators, and show that the ensembles with parameters $L \geq 3$ and $q \geq 2$, and $L = 2$ and $q \geq 3$ are good for the BEC, in the sense that, asymptotically, their typical $h_{\min}$ grows linearly with the block length. In particular, we need to prove that $\bar{s}_{h \leq \hbar}^{\mathcal{C}_{\text{RMA}}}$ tends to zero as $N \longrightarrow \infty$ for all $\hbar < \rho_0 N$, for some value $\rho_0$ between 0 and 1. To prove this, we follow similar arguments to the ones used in [10] and [11, 12] for the asymptotic $d_{\min}$.

Using (3) and (4) in (1), the conditional support size enumerating function (with $w > 0$) of RMA code ensembles can be written as

$$\bar{s}_{w, h_1, \ldots, h_{L-1}, h}^{\mathcal{C}_{\text{RMA}}} = \frac{\binom{K}{w} \sum_{d_1=1}^{\lfloor \frac{qw}{2} \rfloor} \binom{N - h_1}{d_1} \binom{h_1 - 1}{d_1 - 1} \binom{h_1 - d_1}{qw - 2d_1}}{\binom{N}{qw}}$$

$$\times \prod_{l=2}^{L-1} \frac{\sum_{d_l=1}^{\lfloor \frac{h_{l-1}}{2} \rfloor} \binom{N - h_l}{d_l} \binom{h_l - 1}{d_l - 1} \binom{h_l - d_l}{h_{l-1} - 2d_l}}{\binom{N}{h_{l-1}}}$$

$$\times \frac{\sum_{d_L=1}^{\lfloor \frac{h_{L-1}}{2} \rfloor} \binom{N - h}{d_L} \binom{h - 1}{d_L - 1} \binom{h - d_L}{h_{L-1} - 2d_L}}{\binom{N}{h_{L-1}}}$$

$$= \sum_{d_1=1}^{\lfloor \frac{qw}{2} \rfloor} \sum_{d_2=1}^{\lfloor \frac{h_1}{2} \rfloor} \cdots \sum_{d_L=1}^{\lfloor \frac{h_{L-1}}{2} \rfloor} \bar{s}_{w, h_1, \ldots, h_{L-1}, d_1, \ldots, d_L, h}^{\mathcal{C}_{\text{RMA}}}. \tag{10}$$

Without loss of generality, we can write

$$w = \alpha N^a, \ h_i = \beta_i N^{b_i}, \ i = 1, \ldots, L-1,$$
$$h = \rho N^c, \text{ and } d_i = \gamma_i N^{e_i}, \ i = 1, \ldots, L \tag{11}$$

where $0 \leq a \leq b_1 \leq b_2 \leq \cdots \leq b_{L-1} \leq c \leq 1$, $0 \leq e_1 \leq a \leq 1$, and $0 \leq e_i \leq b_{i-1} \leq 1$, $i = 2, \ldots, L$. These inequalities can be derived from the binomial coefficients in the expression in (10) combined with the fact that for a binomial coefficient $\binom{n}{k}$, $n \geq k \geq 0$. Also, $\alpha, \beta_1, \ldots, \beta_{L-1}, \gamma_1, \ldots, \gamma_L$, and $\rho$ are positive constants. We must consider two cases: 1) at least one of the quantities $w, h_1, \ldots, h_{L-1}, d_1, \ldots, d_L$, or $h$ is of order $o(N)$, and 2) all quantities $w, h_1, \ldots, h_{L-1}, d_1, \ldots, d_L$, and $h$ can be expressed as fractions of the block length $N$, i.e., $a = b_1 = \cdots = b_{L-1} = d_1 = \cdots = d_L = c = 1$. The following lemma addresses the first case for RAA code ensembles.

*Lemma 3:* In the ensemble of RAA codes with block length $N$ and $q \geq 3$, in the case where at least one of the quantities $w, h_1, d_1, d_2$, or $h$ is of order $o(N)$, $N^5 \bar{s}_{w, h_1, d_1, d_2, h}^{\mathcal{C}_{\text{RAA}}} \longrightarrow 0$ as $N \longrightarrow \infty$ for all values of $h$, $1 \leq h < N/2$.

*Proof:* See Appendix A. ∎

Lemma 3 can be generalized to the case of RMA codes with $L \geq 3$ (with or without puncturing) and to the HCC ensembles in Section VII. The proofs are omitted for brevity. As a consequence of Lemma 3, the contribution of the first case to $\bar{s}_{h \leq \hbar}^{\mathcal{C}_{\text{RMA}}}$ tends to zero as $N \longrightarrow \infty$, and we can assume that $w, h_1, \ldots, h_{L-1}, d_1, \ldots, d_L$, and $h$ are all linear in the block length.

We now address the second case by deriving an expression for the stopping set size spectral shape function in (6) for RMA code ensembles. Using Stirling's approximation for the binomial coefficient $\binom{n}{k} \overset{n \to \infty}{\longrightarrow} e^{n \mathbb{H}(k/n)}$ where $\mathbb{H}(\cdot)$ is the binary entropy function with natural logarithms, and the fact that $w, h_1, \ldots, h_{L-1}, d_1, \ldots, d_L$, and $h$ are all assumed to be

TABLE I
ASYMPTOTIC $h_{\min}$ GROWTH RATE COEFFICIENTS FOR RMA CODE ENSEMBLES.

|  | $q=2$ | $q=3$ | $q=4$ | $q=5$ | $q=6$ |
| --- | --- | --- | --- | --- | --- |
| $\rho_0$ ($h_{\min}$) ($L=2$) | N/A | 0.0929 | 0.1289 | 0.1505 | 0.1647 |
| $\rho_0$ ($d_{\min}$) ($L=2$) [10] | N/A | 0.1323 | 0.1911 | 0.2286 | 0.2549 |
| $\rho_0$ ($h_{\min}$) ($L=3$) | 0.0681 | 0.1037 | 0.1194 | 0.1279 | 0.1331 |
| $\rho_0$ ($d_{\min}$) ($L=3$) [10] | 0.1034 | 0.1731 | 0.2143 | 0.2428 | 0.2643 |
| $\rho_0$ ($h_{\min}$) ($L=4$) | 0.0549 | 0.0716 | 0.0784 | 0.0817 | 0.0835 |
| GVB | 0.1100 | 0.1740 | 0.2145 | 0.2430 | 0.2644 |

of the same order as the block length $N$, (10) can be written as

$$\bar{s}_{w,h_1,\ldots,h_{L-1},h}^{\mathcal{C}_{\text{RMA}}}= \sum_{d_1=1}^{\lfloor \frac{qw}{2} \rfloor} \cdots \sum_{d_L=1}^{\lfloor \frac{h_{L-1}}{2} \rfloor} \exp\{f(\alpha,\beta_1,\ldots,\beta_{L-1},\gamma_1,\ldots,\gamma_L,\rho)\,N + o(N)\} \quad (12)$$

when $N \longrightarrow \infty$. In (12), $\alpha = \frac{w}{K}$ is the normalized input stopping set size, $\rho = \frac{h}{N}$ is the normalized output stopping set size, $\beta_l = \frac{h_l}{N}$ is the normalized output support set size of constituent code $C_l$, $\gamma_l = \frac{d_l}{N}$, and the function $f(\cdot)$ is given by

$$f(\beta_0,\beta_1,\ldots,\beta_{L-1},\gamma_1,\ldots,\gamma_L,\rho)$$
$$= \frac{\mathbb{H}(\beta_0)}{q} - \sum_{l=1}^{L} \mathbb{H}(\beta_{l-1}) + \sum_{l=1}^{L}(1-\beta_l)\mathbb{H}\left(\frac{\gamma_l}{1-\beta_l}\right)$$
$$+ \sum_{l=1}^{L}\beta_l \mathbb{H}\left(\frac{\gamma_l}{\beta_l}\right) + \sum_{l=1}^{L}(\beta_l-\gamma_l)\mathbb{H}\left(\frac{\beta_{l-1}-2\gamma_l}{\beta_l-\gamma_l}\right) \quad (13)$$

where for conciseness we defined $\beta_0 = \alpha$ and $\beta_L = \rho$. Finally, the stopping set size spectral shape function for RMA code ensembles can be written as [6]

$$r_{\text{s}}^{\mathcal{C}_{\text{RMA}}}(\rho) = \sup_{\substack{0<\beta_0,\ldots,\beta_{L-1}\leq 1 \\ 0<\gamma_1,\ldots,\gamma_L \leq 1}} f(\beta_0,\beta_1,\ldots,\beta_{L-1},\gamma_1,\ldots,\gamma_L,\rho) \quad (14)$$

In deriving (14) from (12), we used the fact that $\max^*(x,y) \triangleq \ln(\exp(x)+\exp(y))$ is approximately equal to $\max(x,y)$ when $x$ and $y$ are large and distinct. Similar comments apply when more than two variables are involved.

From (13) and (14) it can easily be verified that the asymptotic stopping set size spectral shape functions of RMA code ensembles satisfy the recursive relation

$$r_{\text{s}}^{\mathcal{C}_{\text{RMA}(L)}}(\rho) = \sup_{0<u\leq 1}\left[r_{\text{s}}^{\mathcal{C}_{\text{RMA}(L-1)}}(u) + \psi(u,\rho)\right]$$

where $r_{\text{s}}^{\mathcal{C}_{\text{RMA}(l)}}$, $l=L-1,L$, is the asymptotic stopping set size spectral shape function with $l$ accumulators, $r_{\text{s}}^{\mathcal{C}_{\text{RMA}(0)}}$ is defined to be the asymptotic stopping set size spectral shape function of a repeat code, and

$$\psi(u,\rho)$$
$$= \sup_{\max(0,u-\rho)\leq \gamma \leq \min(\rho,1-\rho,u/2)} \left[-\mathbb{H}(u) + \rho \mathbb{H}\left(\frac{\gamma}{\rho}\right) + (1-\rho)\mathbb{H}\left(\frac{\gamma}{1-\rho}\right) + (\rho-\gamma)\mathbb{H}\left(\frac{u-2\gamma}{\rho-\gamma}\right)\right]. \quad (15)$$

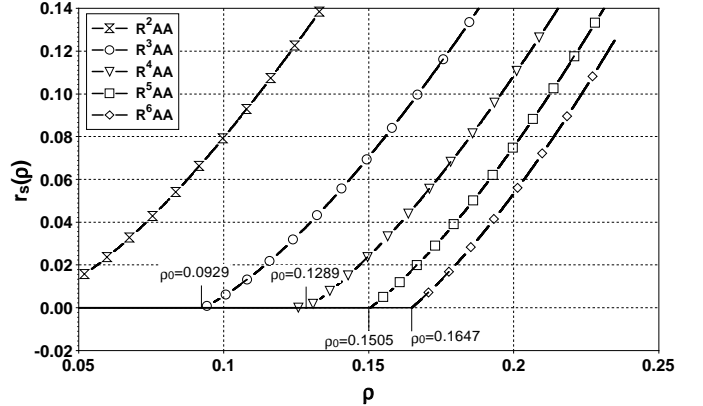

Fig. 6. Asymptotic stopping set size spectral shape function for the RAA code ensemble with $q=2,3,4,5,$ and 6.

To analyze the asymptotic stopping distance behavior of RMA code ensembles, we must solve the optimization problem in (13)-(14). Note that in (14) we did not include the constraints on the relationship of the variables involved in the function $f(\cdot)$, due to lack of space. However, these constraints must be considered in the optimization. The constraints on the involved variables can be derived by looking at the arguments of the binary entropy functions in the expression for the function $f(\cdot)$ in (13). In general, the argument of the binary entropy function should be between 0 and 1. The maximization of the function $f(\cdot)$ is addressed in Appendix B. The numerical evaluation of (13)-(14) is shown in Fig. 6 for RAA code ensembles with $q=2,3,4,5,$ and 6. We observe that the stopping set size spectral shape function for the rate $R=1/2$ RAA code ensemble is strictly positive, meaning that the ensemble is bad for the BEC. For $3 \leq q \leq 6$, the function $r_{\text{s}}^{\mathcal{C}_{\text{RMA}}}(\rho)$ is zero in the range $(0,\rho_0)$ and positive for some values $\rho > \rho_0$. In this case, we cannot conclude directly whether $h_{\min}$ grows linearly with the block length or not. However, we can prove the following theorem, extending the results in [11, 12] to the stopping distance case.

*Theorem 6:* Define $\rho_0 = \max\{\rho^* \in [0,1/2) : r_{\text{s}}^{\mathcal{C}_{\text{RMA}}}(\rho) = 0 \ \forall \rho \leq \rho^*\}$. Assuming that $\lim_{u\rightarrow 0} \frac{\psi(u,\rho)}{u} < 0 \ \forall \rho < \rho_0$, then $\forall \rho^* > 0$

$$\lim_{N\rightarrow \infty} \Pr\left(h_{\min} \leq (\rho_0-\rho^*)N\right) = 0$$

for $L \geq 3$ and $q \geq 2$, and $L=2$ and $q \geq 3$. Thus, if $\rho_0 > 0$ and $r_{\text{s}}^{\mathcal{C}_{\text{RMA}}}(\rho) \geq 0 \ \forall \rho$ (see Lemma 4 in Appendix B), then almost all codes in the ensemble have asymptotic stopping distance growing linearly with $N$ with growth rate coefficient of at least $\rho_0$.

*Proof:* See Appendix C. ∎

We remark that it can be verified that the assumption in Theorem 6 always holds for the numerical values of $\rho_0$ that we have found. From Lemma 3, Theorem 6, and the numerical evaluation of $r_s^{\mathcal{C}_{\mathrm{RAA}}}(\rho)$ in Fig. 6, it results that for RMA codes the typical $h_{\min}$ asymptotically grows linearly with the block length with growth rate of at least $\rho_0$. The exact values of $\rho_0$ are given in Table I. For comparison purposes we also give the asymptotic growth rate coefficient of the $d_{\min}$ computed in [10]. As expected, the asymptotic growth rate coefficient of $h_{\min}$ is smaller than for $d_{\min}$.

TABLE II
ASYMPTOTIC $h_{\min}$ GROWTH RATE COEFFICIENTS FOR PUNCTURED RMA CODE ENSEMBLES WITH $q = 3$ AND NOMINAL CODE RATE $R' = 1/(\lambda q)$.

| $R'$ | 1/3 | 0.35 | 0.37 | 0.38 | 2/5 | 5/12 | 0.43 | 4/9 | 1/2 | 0.54 | 0.55 |
|---|---|---|---|---|---|---|---|---|---|---|---|
| $\lambda$ | 1 | 20/21 | 100/111 | 50/57 | 5/6 | 4/5 | 100/129 | 3/4 | 2/3 | 50/81 | 20/33 |
| $\rho_0$ ($L=2$) | 0.0929 | 0.0911 | 0.0885 | 0.0868 | 0.0820 | 0.0746 | 0.0673 | 0.0585 | 0.0240 | 0.0028 | N/A |
| $\rho_0$ ($L=3$) | 0.1037 | 0.0866 | 0.0632 | 0.0514 | 0.0289 | 0.0124 | 0.0015 | N/A | N/A | N/A | N/A |
| $\rho_0$ ($L=4$) | 0.0716 | 0.0426 | 0.0113 | N/A | N/A | N/A | N/A | N/A | N/A | N/A | N/A |

We can now prove the following theorem.

*Theorem 7:* The typical $h_{\min}$ of RMA code ensembles for $L \geq 3$ and $q \geq 2$ grows linearly with block length.

*Proof:* Note that if we serially concatenate any encoder whose $h_{\min}$ grows linearly with the block length with growth rate coefficient of at least $\rho_0$ with an accumulate code through a uniform interleaver, the resulting concatenated code ensemble will exhibit a $h_{\min}$ growing linearly with the block length with growth rate coefficient of at least $\lceil \rho_0/2 \rceil$. This follows from the fact that the output support set size $h$ of an accumulate code is lower bounded by $\lceil \frac{w}{2} \rceil$ in (3). In more detail, due to the binomial coefficient $\binom{h-d}{w-2d}$, $h - d \geq w - 2d$, from which it follows that $w \leq h + d \leq h + \lfloor w/2 \rfloor$, which implies that $h \geq \lceil \frac{w}{2} \rceil$. Since we know that the RAA code ensemble exhibits a typical $h_{\min}$ that grows linearly with the block length, the theorem is proved for $q \geq 3$. For $q = 2$, it can be shown that the function $r_s^{\mathcal{C}_{\mathrm{RAAA}}}(\rho)$ for the repeat triple-accumulate (RAAA) code ensemble is zero in the range $(0, \rho_0)$, with $\rho_0 > 0$, and positive for some $\rho > \rho_0$. Therefore, by repeating the argument for $q \geq 3$, the theorem is also proved for $q = 2$. ∎

In Table I, we also report the asymptotic growth rate coefficient $\rho_0$ for RAAA and repeat quadruple-accumulate (RAAAA) code ensembles. Interestingly, from Table I it follows that, contrary to the asymptotic $d_{\min}$ growth rate coefficient, which increases with the number of accumulators and tends to approach the GVB [10–12], the asymptotic growth rate coefficient of $h_{\min}$ decreases with the number of accumulators concatenated in series. An intuitive explanation to this behavior can be formulated as follows. From Fig. 3, it follows that there is a many-to-one mapping from input sequences to output sequences in an $N$-fold concatenation of $\mathcal{T}_{\mathrm{ext}}$, in the sense that for a given output sequence there are many input sequences that can produce it. This is not the case for the basic trellis module, i.e., for $\mathcal{T}_{\mathrm{ext}}$ in Fig. 3 without the transition $(1, 1/1, 1)$, where the mapping is one-to-one. In a concatenation of $L$ trellises based on the basic trellis module, there is still a one-to-one mapping between input sequences and output sequences. However, in a concatenation of $L$ trellises based on $\mathcal{T}_{\mathrm{ext}}$, there will be an increasing set (in $L$) of input sequences that map to a particular output sequence. This phenomenon can be easily understood by considering the probabilistic bound (5) on $h_{\min}$ for the finite-length regime. Since the set of input sequences that map to a particular output sequence increases with $L$, the terms $\bar{s}_h^{\mathcal{C}_{\mathrm{RMA}}} = \sum_{w=1}^{K} \bar{s}_{w,h}^{\mathcal{C}_{\mathrm{RMA}}}$ in the right hand side of (5) will also increase. Consequently, $h_{\min}$ will decrease. Our conjecture is that the same phenomenon applies to the asymptotic $h_{\min}$ growth rate coefficient. From these results, it is apparent that serially concatenated codes with more than three encoding stages are not well suited for the BEC.

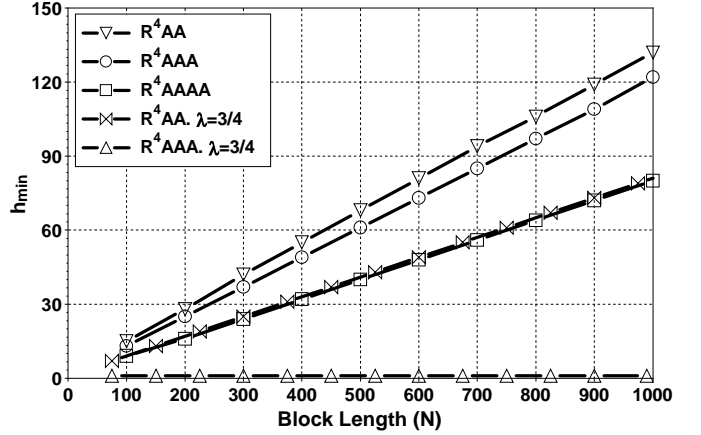

Fig. 7. Probabilistic lower bound on the stopping distance $h_{\min}$ versus block length $N$ for RMA codes with $q = 4$ and $L = 2, 3$, and 4.

In Fig. 7, we plot the probabilistic lower bound on $h_{\min}$ from Lemma 2 for RMA codes with $q = 4$, $L = 2, 3$, and 4, and codeword length $N$ up to 1000 bits. The bounds were obtained by setting $\varepsilon = 0.5$ in (5), i.e., at least half of the codes in the ensemble have stopping distance at least equal to the value indicated by the curves. The results are in agreement with the asymptotic $h_{\min}$ growth rates in Table I. The best growth rate is obtained for $L = 2$, while increasing the number of accumulate codes decreases the growth rate.

### C. RMA Codes with Random Puncturing

In this section, we consider high rate RMA code ensembles obtained by puncturing the output of the most inner accumulator $C_L$. We assume that code bits at the output of $C_L$ are punctured randomly, since otherwise we cannot guarantee asymptotic linear growth rate for $h_{\min}$. Denote by $\lambda$ ($0 \leq \lambda \leq 1$) the puncturing permeability rate, i.e., the fraction of bits surviving after puncturing, and by $R' = R/\lambda = 1/(\lambda q)$ the nominal code rate of the punctured RMA code.

The SIOSEF of a randomly punctured code $C^{\mathrm{punct.}}$ with input support set size $w$, output support set size $h$ before puncturing, and output support set size $h'$ after puncturing

TABLE III
ASYMPTOTIC $h_{\min}$ GROWTH RATE COEFFICIENTS FOR PUNCTURED RMA CODE ENSEMBLES WITH $q = 4$ AND NOMINAL CODE RATE $R' = 1/(\lambda q)$.

| $R'$ | 1/4 | 0.28 | 0.29 | 3/10 | 5/16 | 0.33 | 1/3 | 11/30 | 2/5 | 0.41 | 0.42 | 0.43 |
|---|---|---|---|---|---|---|---|---|---|---|---|---|
| $\lambda$ | 1 | 25/28 | 25/29 | 5/6 | 4/5 | 25/33 | 3/4 | 15/22 | 5/8 | 25/41 | 25/42 | 25/43 |
| $\rho_0$ ($L=2$) | 0.1289 | 0.1192 | 0.1142 | 0.1077 | 0.0977 | 0.0819 | 0.0788 | 0.0474 | 0.0188 | 0.0112 | 0.0045 | N/A |
| $\rho_0$ ($L=3$) | 0.1194 | 0.0694 | 0.0528 | 0.0373 | 0.0198 | 0.0004 | N/A | N/A | N/A | N/A | N/A | N/A |
| $\rho_0$ ($L=4$) | 0.0784 | 0.0112 | N/A | N/A | N/A | N/A | N/A | N/A | N/A | N/A | N/A | N/A |

is given by [10]

$$a_{w,h'}^{C^{\text{punct.}}} = \sum_{h=h'}^{N} a_{w,h}^{C} \frac{\binom{h}{h'}\binom{N-h}{\lambda N - h'}}{\binom{N}{\lambda N}}. \quad (16)$$

Using Stirling's approximation in (16) and a generalization of Lemma 3, and coupling it with (13) and (14), the stopping set size spectral shape function of a punctured RMA code ensemble is given by

$$r_{\text{s}}^{C_{\text{RMA}}^{\text{punct.}}}(\rho') = \frac{1}{\lambda} \sup_{\substack{0 < \beta_0, \ldots, \beta_L \leq 1 \\ 0 < \gamma_1, \ldots, \gamma_L \leq 1}} \frac{\mathbb{H}(\beta_0)}{q} - \sum_{l=1}^{L} \mathbb{H}(\beta_{l-1})$$
$$+ \sum_{l=1}^{L}(1-\beta_l)\mathbb{H}\left(\frac{\gamma_l}{1-\beta_l}\right) + \sum_{l=1}^{L}\beta_l\mathbb{H}\left(\frac{\gamma_l}{\beta_l}\right)$$
$$+ \sum_{l=1}^{L}(\beta_l - \gamma_l)\mathbb{H}\left(\frac{\beta_{l-1} - 2\gamma_l}{\beta_l - \gamma_l}\right) + \beta_L \mathbb{H}\left(\frac{\lambda \rho'}{\beta_L}\right)$$
$$+ (1-\beta_L)\mathbb{H}\left(\frac{\lambda(1-\rho')}{1-\beta_L}\right) - \mathbb{H}(\lambda)$$

where $\rho' = \frac{h'}{\lambda N}$ is the normalized stopping set size after puncturing.

The values of $\rho_0$ corresponding to $r_{\text{s}}^{C_{\text{RMA}}^{\text{punct.}}}(\rho')$ are given in Tables II and III for $L = 2, 3,$ and $4$ mother RMA code ensembles with $q = 3$ and $4$, respectively, for several nominal code rates $R'$. Asymptotic linear growth can be guaranteed for some rates $R' > 1/q$. However, it is interesting to note that the asymptotic stopping set size spectral shape function is strictly positive with heavy puncturing of the mother code ensemble, which implies that the asymptotic linear growth rate property breaks down with heavy puncturing. For instance, for $L = 2$ and $q = 3$, the punctured ensemble remains good for the BEC up to rate $R' = 0.54$, but for heavier puncturing this property is lost. This phenomenon is more significant for punctured RMA code ensembles with a larger number of accumulators and for larger $q$. For the $L = 3$ and $q = 3$ punctured RMA code ensemble, the asymptotic linear growth rate property is lost between $R' = 0.43$ and $R' = 4/9$. In Fig. 7, we plot the probabilistic lower bound on $h_{\min}$ from Lemma 2 for punctured $q = 4$ RMA code ensembles with $\lambda = 3/4$ and $L = 2$ and $3$. The results are in agreement with the asymptotic analysis; for $L = 2$ and permeability rate $\lambda = 3/4$ linear growth rate is guaranteed. However, when applying the same puncturing to the $L = 3$ RMA code ensemble, the asymptotic stopping set size spectral shape function becomes strictly positive, and the linear growth rate property breaks down. Note that these results are in contrast with the results in [10], where it was observed that the asymptotic normalized $d_{\min}$ gets closer to the GVB for higher rates with random puncturing.

## VII. STOPPING DISTANCE ANALYSIS FOR HCCS

In this section, we address the behavior of $h_{\min}$ for HCC ensembles in the form of Fig. 2. For brevity, we only give the conditional support size enumerating function (with $w > 0$) and the stopping set size spectral shape function for type-1 and type-4 HCC ensembles.

### A. Type-1 HCC

We consider the type-1 HCC ensemble in Fig. 2. Let $q$ be the number of accumulators of the outer MPCC and let $q+1$ be the index denoting the inner accumulator. For convenience, we define $h_p = h_1 + \cdots + h_q$, the input support set size of the inner accumulator $C_{q+1}$. Since none of the parallel branches of the outer MPCC are connected to the channel, the set $Q$ is empty. Using (2) and (3), the conditional support size enumerating function (with $w > 0$) for the type-1 HCC ensemble can now be written as

$$\bar{s}_{w,h_1,\ldots,h_q,h}^{C_{\text{t1}}} = \frac{\prod_{l=1}^{q} \sum_{d_l=1}^{\lfloor \frac{w}{2} \rfloor} \binom{K-h_l}{d_l}\binom{h_l-1}{d_l-1}\binom{h_l-d_l}{w-2d_l}}{\binom{K}{w}^{q-1}}$$
$$\times \frac{\sum_{d_{q+1}=1}^{\lfloor \frac{h_p}{2} \rfloor} \binom{N-h}{d_{q+1}}\binom{h-1}{d_{q+1}-1}\binom{h-d_{q+1}}{h_p - 2d_{q+1}}}{\binom{N}{h_p}}.$$

Now, let $\alpha = \frac{w}{K}$, $\rho = \frac{h}{N}$, $\beta_l = \frac{h_l}{K}$, and $\gamma_l = \frac{d_l}{K}$, where $l = 1, \ldots, q$. We also denote the normalized input support set size of the inner accumulator by $\beta_p = \frac{h_p}{N}$ and define $\delta = \frac{d_{q+1}}{N}$. Using Stirling's approximation and a generalization of Lemma 3, we get

$$r_{\text{s}}^{C_{\text{t1}}}(\rho) = \sup_{\substack{0 < \alpha, \beta_1, \ldots, \beta_q \leq 1 \\ 0 < \delta, \gamma_1, \ldots, \gamma_q \leq 1}} \frac{1}{q}\sum_{l=1}^{q}(1-\beta_l)\mathbb{H}\left(\frac{\gamma_l}{1-\beta_l}\right)$$
$$+ \frac{1}{q}\sum_{l=1}^{q}\beta_l \mathbb{H}\left(\frac{\gamma_l}{\beta_l}\right) + \frac{1}{q}\sum_{l=1}^{q}(\beta_l - \gamma_l)\mathbb{H}\left(\frac{\alpha - 2\gamma_l}{\beta_l - \gamma_l}\right)$$
$$+ (1-\rho)\mathbb{H}\left(\frac{\delta}{1-\rho}\right) + \rho \mathbb{H}\left(\frac{\delta}{\rho}\right)$$
$$+ (\rho - \delta)\mathbb{H}\left(\frac{\beta_p - 2\delta}{\rho - \delta}\right) - \frac{q-1}{q}\mathbb{H}(\alpha) - \mathbb{H}(\beta_p). \quad (17)$$

The numerical evaluation of (17) (the details are omitted due to lack of space) is given in Fig. 8 for $q = 4$. We remark that the optimization is harder than for the $d_{\min}$ case, since the objective function involves more variables. We observe that $r_{\text{s}}^{C_{\text{t1}}}(\rho)$ is strictly zero in the range $(0, \rho_0 = 0.1289)$ and positive for some $\rho > \rho_0$. From these results and the generalizations of Lemma 3 and Theorem 6 to HCC ensembles, $\rho_0 = 0.1289$ is a lower bound on the asymptotic $h_{\min}$ growth rate coefficient of the ensemble. It is worth mentioning that the supremum is obtained for $\beta_1 = \cdots = \beta_q$ and $\gamma_1 = \cdots = \gamma_q$, i.e., when all the accumulators of the outer MPCC contribute equally to the stopping set size. Therefore, we can set $\beta_1 = \cdots = \beta_q = \beta_p$ and $\gamma_1 = \cdots = \gamma_q$ in (17), and the expression for the stopping set size spectral shape function becomes equal to (13) (with $L = 2$) for RAA code ensembles. Indeed, the type-1 HCC ensemble and the RAA code ensemble are identical in both asymptotic $h_{\min}$ growth rate and convergence threshold (see Section VIII below).

### B. Type-4 HCC

We consider the type-4 HCC ensemble formed by an outer MPCC with $q$ constituent encoders followed by an accumulator. The first encoder (the systematic branch) performs an identity mapping and is sent straight through the channel (therefore $Q = \{1\}$ in (2)). Denote the input support set size of the inner accumulator by $h_p = h_2 + \cdots + h_q$. Also, let $M = (q-1)K = \frac{q-1}{q}N$. Using (2) and (3), the conditional support size enumerating function (with $w > 0$) of the type-4 HCC ensemble can be written as

$$\bar{s}^{C_{t4}}_{w,h_2,\ldots,h_q,h} = \frac{\prod_{l=2}^{q} \sum_{d_l=1}^{\lfloor w/2 \rfloor} \binom{K-h_l}{d_l}\binom{h_l-1}{d_l-1}\binom{h_l-d_l}{w-2d_l}}{\binom{K}{w}^{q-2}}$$
$$\times \frac{\sum_{d_{q+1}=1}^{\lfloor h_p/2 \rfloor} \binom{M-h+w}{d_{q+1}}\binom{h-w-1}{d_{q+1}-1}\binom{h-w-d_{q+1}}{h_p-2d_{q+1}}}{\binom{M}{h_p}}.$$

Again, we define $\alpha = \frac{w}{K}$, $\rho = \frac{h}{N}$, $\beta_l = \frac{h_l}{K}$, and $\gamma_l = \frac{d_l}{K}$, where $l = 2, \ldots, q$. Also, define the normalized size $\beta_p = \frac{h_p}{M}$ and $\delta = \frac{d_{q+1}}{M}$. Using Stirling's approximation and a generalization of Lemma 3, the stopping set size spectral shape function of the type-4 HCC ensemble can now be written as

$$r_s^{C_{t4}}(\rho) = \sup_{\substack{0<\alpha,\beta_2,\ldots,\beta_q\leq 1 \\ 0<\delta,\gamma_2,\ldots,\gamma_q\leq 1}} \frac{1}{q}\sum_{l=2}^{q}(1-\beta_l)\mathbb{H}\left(\frac{\gamma_l}{1-\beta_l}\right)$$
$$+ \frac{1}{q}\sum_{l=2}^{q}\beta_l\mathbb{H}\left(\frac{\gamma_l}{\beta_l}\right) + \frac{1}{q}\sum_{l=2}^{q}(\beta_l-\gamma_l)\mathbb{H}\left(\frac{\alpha-2\gamma_l}{\beta_l-\gamma_l}\right)$$
$$+ \left(\frac{q-1+\alpha}{q}-\rho\right)\mathbb{H}\left(\frac{(q-1)\delta}{q-1-q\rho+\alpha}\right)$$
$$+ \left(\rho-\frac{\alpha}{q}\right)\mathbb{H}\left(\frac{(q-1)\delta}{q\rho-\alpha}\right)$$
$$+ \left(\rho-\frac{\alpha+(q-1)\delta}{q}\right)\mathbb{H}\left(\frac{(q-1)(\beta_p-2\delta)}{q\rho-\alpha-(q-1)\delta}\right)$$
$$- \frac{q-2}{q}\mathbb{H}(\alpha) - \frac{q-1}{q}\mathbb{H}(\beta_p).$$

The stopping set size spectral shape functions for type-2 and type-3 HCCs can be obtained in a similar manner. The stopping set size spectral shape functions for the type-2, type-3, and type-4 HCC ensembles are also plotted in Fig. 8 for $q = 4$ ($q_1 = 1$ for the type-2 HCC ensemble). The largest asymptotic $h_{\min}$ growth rate ($\rho_0 = 0.1289$) is obtained for the type-1 HCC ensemble. If one of the accumulators of the outer MPCC is replaced by a feedforward branch (type-2 HCC ensemble), the asymptotic $h_{\min}$ growth rate is decreased to $\rho_0 = 0.1207$. The asymptotic $h_{\min}$ growth rate is further decreased to $\rho_0 = 0.0886$ if the feedforward branch does not enter the inner accumulator. Finally, the smallest asymptotic $h_{\min}$ growth rate ($\rho_0 = 0.0829$) is obtained for the fully systematic type-4 HCC ensemble. As a comparison, the asymptotic growth rate coefficient of the $d_{\min}$ computed in [8] is 0.1911, 0.1793, 0.1350, and 0.1179 for type-1, type-2, type-3, and type-4 HCCs, respectively. The same behavior is observed for other values of $q$. The coefficients $\rho_0$ for the four HCC ensembles with $q = 3, 4, 5$, and $6$ are given in Table IV.

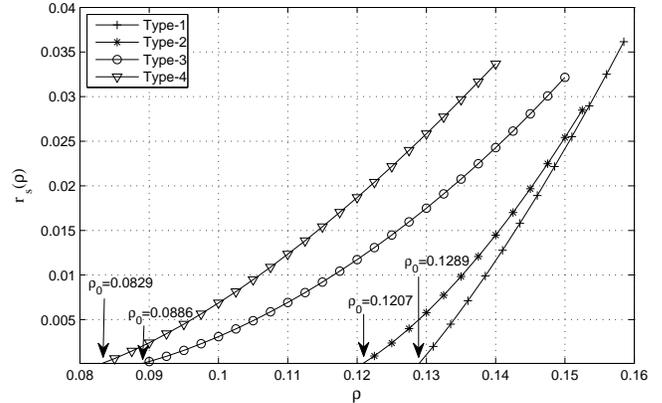

Fig. 8. Asymptotic stopping set size spectral shape function for the rate-1/4 HCC ensembles depicted in Fig. 2.

TABLE IV
ASYMPTOTIC $h_{\min}$ GROWTH RATE COEFFICIENTS FOR HCC ENSEMBLES.

| $\rho_0$ | $q=3$ | $q=4$ | $q=5$ | $q=6$ |
|---|---|---|---|---|
| Type-1 | 0.0929 | 0.1289 | 0.1505 | 0.1647 |
| Type-2 | 0.0716 | 0.1207 | 0.1462 | 0.1623 |
| Type-3 | N/A | 0.0886 | 0.1266 | 0.1494 |
| Type-4 | N/A | 0.0829 | 0.1199 | 0.1430 |

## VIII. EXIT CHARTS ANALYSIS

In the previous sections, we have shown that RMA and HCC ensembles are good for the BEC. In this section, we address iterative constituent code oriented decoding of RMA codes and HCCs on the BEC by using EXIT charts analysis [29] to estimate their convergence thresholds. In particular, we follow the definitions of the EXIT functions in [30].

Denote by $\mathbf{u} = (u_1, \ldots, u_K)$ the sequence of information symbols which is mapped to the sequence of code symbols $\mathbf{x} = (x_1, \ldots, x_N)$ by an $(N, K)$ code $C$. As required for the EXIT charts analysis, we assume that $N \longrightarrow \infty$. The information symbols are transmitted over a BEC, called the information-symbol *a priori* channel, with erasure probability $p_{a_u}$. For information symbol $u_i$, the received symbol is denoted by $y_{u_i}$, and the corresponding *a priori* L-value (or log-likelihood ratio) by $L_a^C(u_i)$. Also, let $I(u_i; L_a^C(u_i))$ be the mutual information (MI) between $u_i$ and $L_a^C(u_i)$. The average *a priori* MI for the information symbols is

$$I_{a,u}^C = \frac{1}{K}\sum_{i=1}^{K} I\left(u_i; L_a^C(u_i)\right) = 1 - p_{a_u}.$$

Similarly, the code symbols are transmitted over a BEC, called the code-symbol *a priori* channel, with erasure probability $p_{a_x}$. For code symbol $x_i$, the received symbol is denoted by $y_{x_i}$, and the corresponding *a priori* L-value by $L_a^C(x_i)$. Also, let $I(x_i; L_a^C(x_i))$ be the MI between $x_i$ and $L_a^C(x_i)$. The average *a priori* MI for the code symbols is

$$I_{a,x}^C = \frac{1}{N} \sum_{i=1}^{N} I\left(x_i; L_a^C(x_i)\right) = 1 - p_{a_x}.$$

The *a priori* L-values $L_a^C(u_i)$ and $L_a^C(x_i)$ are given to an *a posteriori* probability (APP) decoder, which computes the extrinsic L-values $L_e^C(u_i)$ and $L_e^C(x_i)$ for information symbols and code symbols, respectively. The average extrinsic MI for information and code symbols are

$$I_{e,u}^C = \frac{1}{K} \sum_{i=1}^{K} I\left(u_i; L_e^C(u_i)\right)$$

and

$$I_{e,x}^C = \frac{1}{N} \sum_{i=1}^{N} I\left(x_i; L_e^C(x_i)\right)$$

respectively. The input-output behavior of an APP decoder is then completely characterized by two EXIT functions $T_u$ and $T_x$ which specify the evolution of the extrinsic MIs as a function of the *a priori* MIs. In particular, we can write $I_{e,u}^C = T_u(I_{a,u}^C, I_{a,x}^C)$ and $I_{e,x}^C = T_x(I_{a,u}^C, I_{a,x}^C)$. A nice property of the BEC is that the EXIT functions for the repeat code and for convolutional encoders can be given in closed form as shown in [30] and [31].

In a concatenated coding scheme consisting of $P$ constituent encoders, decoding proceeds by alternating between the corresponding $P$ constituent decoders. The iterative decoding process can then be tracked using a multi-dimensional EXIT chart [32]. Alternatively, the EXIT functions of the constituent decoders can be properly combined and projected into a two-dimensional chart [33]. For instance, convergence of the RMA and HCC ensembles can be visualized using a two-dimensional EXIT chart reporting in a single figure the EXIT function of the outer code (the concatenation of a repeat code and $L - 1$ accumulators for the RMA code ensemble, and the MPCC in the case of HCCs) and the EXIT function of the inner accumulator. Consider as an example the type-1 and type-2 HCCs of Fig. 2. As defined above, let $L_a^{C_l}(u_i^l)$ and $L_a^{C_l}(x_i^l)$ (resp. $L_e^{C_l}(u_i^l)$ and $L_e^{C_l}(x_i^l)$) be the *a priori* (resp. extrinsic) L-values for the information and code symbols of constituent encoder $C_l$, respectively. Also, let $I_{a,u^l}^{C_l}$ and $I_{a,x^l}^{C_l}$ (resp. $I_{e,u^l}^{C_l}$ and $I_{e,x^l}^{C_l}$) be the corresponding MIs. The EXIT functions for the constituent decoders of the MPCC (i.e., $l = 1, \ldots, q$) can be expressed as

$$I_{e,u^l}^{C_l} = T_u^{C_l}\left(\boxplus_{i=1, i \neq l}^{q} I_{e,u^i}^{C_i}, I_{a,x^l}^{C_l}\right)$$

$$I_{e,x^l}^{C_l} = T_x^{C_l}\left(\boxplus_{i=1, i \neq l}^{q} I_{e,u^i}^{C_i}, I_{a,x^l}^{C_l}\right)$$

where $I_1 \boxplus \ldots \boxplus I_n = 1 - (1 - I_1) \cdots (1 - I_n)$ [34]. Note that $I_{a,x^l}^{C_l} = I_{a,x^{\text{MPCC}}}^{C_{\text{MPCC}}} = 1 - p_{a,x^{\text{MPCC}}}$. The EXIT function of the outer MPCC $I_{e,x^{\text{MPCC}}}^{C_{\text{MPCC}}}$ can be computed for all values $0 \leq I_{a,x^{\text{MPCC}}}^{C_{\text{MPCC}}} \leq 1$ by activating all $q$ decoders of the MPCC until $I_{e,u^l}^{C_l}$ and $I_{e,x^l}^{C_l}$ converge to a fixed value, and setting $I_{e,x^{\text{MPCC}}}^{C_{\text{MPCC}}} = \frac{1}{q} \sum_{l=1}^{q} I_{e,x^l}^{C_l}$. Finally, the EXIT chart of the HCC is obtained by reporting in a single plot the functions $I_{e,x^{\text{MPCC}}}^{C_{\text{MPCC}}}(0, I_{a,x^{\text{MPCC}}}^{C_{\text{MPCC}}})$ and $I_{e,u^{q+1}}^{C_{q+1}}(I_{a,u^{q+1}}^{C_{q+1}}, I_{a,x^{q+1}}^{C_{q+1}})$, where $I_{a,x^{\text{MPCC}}}^{C_{\text{MPCC}}} = I_{e,u^{q+1}}^{C_{q+1}}$ and $I_{a,x^{q+1}}^{C_{q+1}} = 1 - p_{\text{ch}}$, where $p_{\text{ch}}$ is the erasure probability of the communication channel. The EXIT charts for RMA code ensembles and type-3 and type-4 HCCs can be obtained using a similar procedure. The convergence thresholds, i.e., the largest values of the channel erasure probability $p_{\text{ch}}$ such that there is an open tunnel in the EXIT charts, for RMA and HCC ensembles for $q = 3, \ldots, 6$ are given in Table V. From Tables V and IV the presence of a tradeoff between asymptotic $h_{\min}$ growth rate and convergence threshold for HCC ensembles can be observed. In fact, the hierarchy arising from Table IV is completely reversed in Table V. The type-1 code ensemble is the best one in terms of asymptotic $h_{\min}$ growth rate. However, it has the worst convergence among the four considered HCCs. The convergence threshold can be significantly improved if one of the parallel branches of the outer MPCC is sent straight through the channel, at the expense of a smaller asymptotic $h_{\min}$ growth rate coefficient. The best convergence is achieved by the fully systematic type-4 ensemble. On the other hand, RAAA and RAAAA code ensembles show very poor thresholds, which make them impractical.

TABLE V
CONVERGENCE THRESHOLDS FOR RMA AND HCC ENSEMBLES.

|  | $q = 3$ | $q = 4$ | $q = 5$ | $q = 6$ |
|---|---|---|---|---|
| Type-1/RAA | 0.4965 | 0.5422 | 0.5719 | 0.5935 |
| Type-2 | 0.5058 | 0.5543 | 0.5847 | 0.6062 |
| Type-3 | 0.5624 | 0.6008 | 0.6252 | 0.6429 |
| Type-4 | 0.6007 | 0.6373 | 0.6582 | 0.6730 |
| RAAA | 0.3259 | 0.3531 | 0.3718 | 0.3860 |
| RAAAA | 0.1957 | 0.2105 | 0.2209 | 0.2290 |

From the EXIT charts analysis and the asymptotic analysis in Sections VI and VII, it arises that double serially concatenated code ensembles and HCC ensembles are good ensembles for the BEC, since they provide both high asymptotic $h_{\min}$ growth rates and good convergence behavior, while adding more encoding stages in RMA codes penalizes both the asymptotic $h_{\min}$ growth rate and the convergence threshold.

## IX. CONCLUSION

In this paper, we extended the results of [10] and [11, 12], where RMA code ensembles were proved to be asymptotically good, in the sense that their typical $d_{\min}$ asymptotically grows linearly with the block length, to show that these ensembles are also good for the BEC, i.e., their typical $h_{\min}$ also grows linearly with the block length. However, contrary to the asymptotic $d_{\min}$, whose growth rate coefficient increases with the number of accumulators, the growth rate of $h_{\min}$ decreases with the number of encoding stages. Therefore, double serially concatenated codes seem to be good for the BEC, while adding more encoding stages degrades performance. Furthermore, we considered random puncturing of the RMA code ensemble to achieve higher rates and showed that the asymptotic stopping set size spectral shape function is strictly positive for high rates, which implies that the asymptotic linear growth rate property of $h_{\min}$ is lost when the rate is high. In particular, for each RMA code ensemble, i.e., for each pair $L$ and $q$, there is a particular code rate for which the linear growth rate property breaks down. This phenomenon becomes more significant for larger values of $L$ and $q$.

We also considered the HCC ensembles recently discussed in [8], and showed that they also exhibit an asymptotic $h_{\min}$ linear growth. A fundamental tradeoff between asymptotic $h_{\min}$ growth rate and convergence threshold was observed for these ensembles. In that sense, HCC ensembles offer more degrees of freedom for code construction than RMA code ensembles and therefore they allow for constructing codes with a better tradeoff between asymptotic $h_{\min}$ growth rate and iterative convergence properties.

## APPENDIX A
## PROOF OF LEMMA 3

From (11) with $L = 2$ we can write
$$w = \alpha N^a, \; h_1 = \beta_1 N^{b_1}, \; h = \rho N^c,$$
$$d_1 = \gamma_1 N^{e_1}, \text{ and } d_2 = \gamma_2 N^{e_2}$$

where $0 \leq a \leq b_1 \leq c \leq 1$, $0 \leq e_1 \leq a \leq 1$, and $0 \leq e_2 \leq b_1 \leq 1$, and $a, b_1, c, e_1,$ and $e_2$ are not all equal to one. Also, $\alpha, \beta_1, \gamma_1, \gamma_2,$ and $\rho$ are positive constants. Now,

$$\binom{K}{w} \frac{\binom{N-h_1}{d_1}\binom{h_1-1}{d_1-1}\binom{h_1-d_1}{qw-2d_1}}{\binom{N}{qw}} \leq \left(\frac{N}{q\alpha N^a}\right)^{\alpha N^a}$$
$$\times \left(\frac{N - \beta_1 N^{b_1}}{\gamma_1 N^{e_1}}\right)^{\gamma_1 N^{e_1}} \left(\frac{\beta_1 N^{b_1}}{\gamma_1 N^{e_1}}\right)^{\gamma_1 N^{e_1}} \left(\frac{N}{q\alpha N^a}\right)^{-q\alpha N^a}$$
$$\times \left(\frac{\beta_1 N^{b_1} - \gamma_1 N^{e_1}}{q\alpha N^a - 2\gamma_1 N^{e_1}}\right)^{q\alpha N^a - 2\gamma_1 N^{e_1}} \exp\left(o(N^a \ln N)\right)$$
(18)

and

$$\frac{\binom{N-h}{d_2}\binom{h-1}{d_2-1}\binom{h-d_2}{h_1-2d_2}}{\binom{N}{h_1}} \leq \left(\frac{N - \rho N^c}{\gamma_2 N^{e_2}}\right)^{\gamma_2 N^{e_2}}$$
$$\times \left(\frac{\rho N^c}{\gamma_2 N^{e_2}}\right)^{\gamma_2 N^{e_2}} \left(\frac{\rho N^c - \gamma_2 N^{e_2}}{\beta_1 N^{b_1} - 2\gamma_2 N^{e_2}}\right)^{\beta_1 N^{b_1} - 2\gamma_2 N^{e_2}}$$
$$\times \left(\frac{N}{\beta_1 N^{b_1}}\right)^{-\beta_1 N^{b_1}} \exp\left(o(N^{b_1} \ln N)\right)$$
(19)

where we have used the inequalities

$$\left(\frac{N}{l}\right)^l \frac{1}{\varphi_N(l-1)} \leq \binom{N}{l} \leq \left(\frac{N}{l}\right)^l \varphi_l(l-1)$$

from [10] to bound the binomial coefficients of the left hand sides of (18) and (19). The rest of the proof is similar to the proof of Lemma 1 in [10], where five cases are considered. Here, we will only consider the first two cases in [10, Eq. (43)] and [10, Eq. (44)] due to lack of space. The other three cases can be proved in a similar fashion.

- Assume $0 \leq a = b_1 \leq c < 1$ ([10, Eq. (43)]) or $0 \leq a = b_1 < c \leq 1$ ([10, Eq. (44)]). In both cases, $a = b_1 < 1$. Define $\gamma_1^* = \gamma_1$ if $a = e_1$ and 0 otherwise. In both cases, $\gamma_1^* \leq \frac{q\alpha}{2}$, since the summation index $d_1$ in (10) is upper-bounded by $\lfloor \frac{qw}{2} \rfloor$, from which it follows that $\gamma_1$ is upper-bounded by $\frac{q\alpha}{2}$ when $a = e_1$. From (18), it follows that

as $N \longrightarrow \infty$, the right hand side of (18) approaches
$$\exp\left((1-a)(\alpha - q\alpha + \gamma_1^*)N^a \ln N + o(N^a \ln N)\right)$$
$$\leq \exp\left((1-a)\alpha \left(1 - \frac{q}{2}\right) N^a \ln N + o(N^a \ln N)\right)$$
$$\longrightarrow 0 \text{ as } N \to \infty$$

since we have assumed that $q \geq 3$. Define $\gamma_2^* = \gamma_2$ if $e_2 = b_1 \,(= a)$ and 0 otherwise. In both cases, $\gamma_2^* \leq \frac{\beta_1}{2}$, since the summation index $d_2$ in (10) is upper-bounded by $\lfloor \frac{h_1}{2} \rfloor$, from which it follows that $\gamma_2$ is upper-bounded by $\frac{\beta_1}{2}$ when $e_2 = b_1 \,(= a)$. From (19), and when $c < 1$, it follows that as $N \longrightarrow \infty$, the right hand side of (19) approaches
$$\exp\left((1-c)(\gamma_2^* - \beta_1)N^a \ln N + o(N^a \ln N)\right)$$
$$\leq \exp\left(-(1-c)\frac{\beta_1}{2} N^a \ln N + o(N^a \ln N)\right)$$
$$\longrightarrow 0 \text{ as } N \longrightarrow \infty.$$

From (19), and when $c = 1$, it follows that as $N \longrightarrow \infty$, the right hand side of (19) approaches $\exp\left(o(N^a \ln N)\right)$, and the result of Lemma 3 is proved for the special cases of $0 \leq a = b_1 \leq c < 1$ and $0 \leq a = b_1 < c \leq 1$.

## APPENDIX B
## MAXIMIZATION OF THE FUNCTION $f(\cdot)$ FOR RMA CODE ENSEMBLES

We consider the maximization of the function $f(\beta_0, \beta_1, \ldots, \beta_{L-1}, \gamma_1, \ldots, \gamma_L, \rho)$ in (13). A maximum can occur on the boundary or in the region $0 < \beta_l, \gamma_l \leq 1 \; \forall l$. The following lemma holds.

*Lemma 4:* The asymptotic stopping set size spectral shape function of the RMA code ensemble is non-negative, i.e.,
$$r_{\text{s}}^{\mathcal{C}_{\text{RMA}}}(\rho) \geq 0, \; \forall \rho \in [0,1].$$

*Proof:* The values of $\gamma_l$ in (14) can be chosen such that $r_{\text{s}}^{\mathcal{C}_{\text{RMA}}}(\rho)$ reduces to the spectral shape function of the code ensemble. Then, we can use [11, Proposition 12], and the result follows. ∎

We consider first the case where the maximum occurs in the region $0 < \beta_l, \gamma_l \leq 1 \; \forall l$. In this case the maximum is attained at the point where all partial derivatives $\frac{\partial f}{\partial \beta_l}$ and $\frac{\partial f}{\partial \gamma_l}$ equal zero. Setting $\frac{\partial f}{\partial \beta_0} = \frac{\partial f}{\partial \gamma_l} = \frac{\partial f}{\partial \beta_l} = 0$ gives

$$\gamma_1 = \frac{\left(1 + \left(\frac{1-\beta_0}{\beta_0}\right)^{\frac{q-1}{q}}\right)\beta_0 - \beta_1}{\left(1 + 2\left(\frac{1-\beta_0}{\beta_0}\right)^{\frac{q-1}{q}}\right)} = A(\beta_0) - B(\beta_0)\beta_1$$
(20)

$$\beta_l = \frac{(1-\gamma_l)(\beta_{l-1} - 2\gamma_l)^2 - \gamma_l^2(\gamma_l - \beta_{l-1})}{\gamma_l^2 + (\beta_{l-1} - 2\gamma_l)^2}, \; l = 1, \ldots, L$$
(21)

$$\gamma_{l+1} =$$
$$\frac{\beta_l^2(\beta_l - \beta_{l+1})(1 - \beta_l - \gamma_l) + \beta_l(1-\beta_l)^2(\beta_l + \gamma_l - \beta_{l-1})}{\beta_l^2(1 - \beta_l - \gamma_l) + 2(1-\beta_l)^2(\beta_l + \gamma_l - \beta_{l-1})}$$
$$= C^{(0)}(\beta_{l-1}, \beta_l, \gamma_l) - D^{(0)}(\beta_{l-1}, \beta_l, \gamma_l)\beta_{l+1},$$
$$l = 1, \ldots, L-1. \quad (22)$$

To determine a solution to the above set of equations, we choose the following strategy. First treat $\beta_0$ as a free parameter and compute $A(\beta_0)$ and $B(\beta_0)$. Then, set $\gamma_1 = A(\beta_0) - B(\beta_0)\beta_1$ in (21) to obtain $\beta_1$ by solving the resulting third order equation. $\gamma_1$ can now be obtained using (20). We are now in the position to compute $C^{(0)}(\beta_0, \beta_1, \gamma_1)$ and $D^{(0)}(\beta_0, \beta_1, \gamma_1)$ in (22) and combine (22) and (21) to obtain the values for $\gamma_2$ and $\beta_2$. Finally, the values for $\gamma_l$ and $\beta_l$ with $l > 2$ can be obtained by using (22) and (21) recursively.

We must also consider the following boundary conditions when they apply.

- The boundary condition $\gamma_l = 1 - \beta_l$ and the condition $\frac{\partial f}{\partial \beta_l} + \frac{\partial f}{\partial \gamma_l} \cdot \frac{\partial \gamma_l}{\partial \beta_l} = 0$ result in

$$\gamma_{l+1} = \frac{(\beta_l - \beta_{l+1})\beta_l^2 + \beta_l(\beta_{l-1} - 2(1-\beta_l))^2}{\beta_l^2 + 2(\beta_{l-1} - 2(1-\beta_l))^2} \quad (23)$$
$$= C^{(1)}(\beta_{l-1}, \beta_l) - D^{(1)}(\beta_{l-1}, \beta_l)\beta_{l+1}.$$

- The boundary condition $\gamma_l = \beta_l$ and the condition $\frac{\partial f}{\partial \beta_l} + \frac{\partial f}{\partial \gamma_l} \cdot \frac{\partial \gamma_l}{\partial \beta_l} = 0$ result in

$$\gamma_{l+1} = \frac{(1-2\beta_l)^2(\beta_l - \beta_{l+1}) + \beta_l(1-\beta_l)^2}{(1-2\beta_l)^2 + 2(1-\beta_l)^2} \quad (24)$$
$$= C^{(2)}(\beta_{l-1}, \beta_l) - D^{(2)}(\beta_{l-1}, \beta_l)\beta_{l+1}.$$

- The boundary condition $\gamma_l = \beta_{l-1} - \beta_l$ and the condition $\frac{\partial f}{\partial \beta_l} + \frac{\partial f}{\partial \gamma_l} \cdot \frac{\partial \gamma_l}{\partial \beta_l} = 0$ result in

$$\gamma_{l+1} =$$
$$\frac{\beta_l^2(\beta_l - \beta_{l+1})(\beta_{l-1} - \beta_l)^2 + \beta_l(2\beta_l - \beta_{l-1})^2(1-\beta_l)^2}{\beta_l^2(\beta_{l-1} - \beta_l)^2 + 2(2\beta_l - \beta_{l-1})^2(1-\beta_l)^2}$$
$$= C^{(3)}(\beta_{l-1}, \beta_l) - D^{(3)}(\beta_{l-1}, \beta_l)\beta_{l+1}.$$

- The boundary condition $\gamma_l = \frac{\beta_{l-1}}{2}$ and the condition $\frac{\partial f}{\partial \beta_l} + \frac{\partial f}{\partial \gamma_l} \cdot \frac{\partial \gamma_l}{\partial \beta_l} = 0$ result in

$$\gamma_{l+1} =$$
$$\frac{\beta_l^2(\beta_l - \beta_{l+1})(1-\beta_l - \frac{\beta_{l-1}}{2}) + \beta_l(1-\beta_l)^2(\beta_l - \frac{\beta_{l-1}}{2})}{\beta_l^2(1-\beta_l - \frac{\beta_{l-1}}{2}) + 2(1-\beta_l)^2(\beta_l - \frac{\beta_{l-1}}{2})}$$
$$= C^{(0)}(\beta_{l-1}, \beta_l, \gamma_l) - D^{(0)}(\beta_{l-1}, \beta_l, \gamma_l)\beta_{l+1}.$$

Note that for each boundary condition, $l$ can vary between 1 and $L$ for the first condition and between 1 and $L-1$ for the second condition. For instance, when the boundary condition $\gamma_l = 1 - \beta_l$, $l = 1, \ldots, L$, is applied, then we substitute the two equations in (21) and (22) with $\gamma_l = 1 - \beta_l$ and the equation in (23). Similarly, when the boundary condition $\gamma_l = \beta_l$, $l = 1, \ldots, L$, is applied, then we substitute the two equations in (21) and (22) with $\gamma_l = \beta_l$ and the equation in (24). In general, all possible combinations must be checked, i.e., applying one of the four boundary conditions or no boundary condition at all when computing the pair $(\beta_l, \gamma_l)$, $l = 1, \ldots, L$, where $\beta_L = \rho$. Using the procedure described above, we can compute the asymptotic $h_{\min}$ growth rate coefficient $\rho_0$ for RMA code ensembles with different values of $q$ and $L$, where $\rho_0$ has been defined in Theorem 6.

## APPENDIX C
## PROOF OF THEOREM 6

The proof of Theorem 6 follows closely the proof of Theorem 11 in [11], which is inspired by asymptotic techniques devised in [35]. The proof of [11, Theorem 11] relies on Lemmas 26 and 27 in [11]. We start by proving [11, Lemma 26] for the stopping set case, which is stated below in our notation for convenience. The lemma is proven by induction on $L$ following the same arguments as in the proof for the codeword case outlined in [12].

*Lemma 5:* Let $\{h_N\}_{N \in \mathbb{N}}$ be a sequence of integers such that for any arbitrary $\eta > 0$

$$\lim_{N \to \infty} \frac{h_N}{N^\eta} = 0 \text{ and } \lim_{N \to \infty} \frac{\ln h_N}{h_N} = 0.$$

Then,

$$\sum_{h=1}^{h_N} \bar{s}_h^{\mathcal{C}_{\text{RMA}}} = O\left(N^{1 - \sum_{i=1}^{L} \lceil \frac{q}{2^i} \rceil + \eta}\right)$$

where $L$ is the number of accumulators.

*Proof:* We prove the lemma by induction on the number of accumulators $L$. Consider first the case of $L = 1$. We have

$$\sum_{h=1}^{h_N} \bar{s}_h^{\mathcal{C}_{\text{RA}}} = \sum_{w=1}^{2h_N/q} \binom{N/q}{w} \sum_{h=1}^{h_N} \frac{\sum_{d=1}^{\lfloor \frac{qw}{2} \rfloor} \binom{N-h}{d}\binom{h-1}{d-1}\binom{h-d}{qw-2d}}{\binom{N}{qw}}$$
$$\leq \sum_{w=1}^{2h_N/q} N^{w - \lceil \frac{qw}{2} \rceil} g(w, N) \sum_{h=1}^{h_N} h^{qw + \lfloor \frac{qw}{2} \rfloor - 3}$$

where (see (7))

$$g(w, N) = \frac{(qw)! e^{qw+w-1}\varphi_N(qw-1)}{q^w w^w} \sum_{d=1}^{\lfloor \frac{qw}{2} \rfloor} \frac{(qw-2d)^{2d-qw}}{d^d(d-1)^{d-1}}.$$

Note that the upper bound of $2h_N/q$ in the summation over $w$ is due to the binomial $\binom{h-d}{qw-2d}$. In more detail, $h-d \geq qw - 2d$, from which it follows that $qw \leq h + d \leq h + \lfloor qw/2 \rfloor$, which implies that $w \leq 2h/q$. Now, it follows that

$$\sum_{h=1}^{h_N} \bar{s}_h^{\mathcal{C}_{\text{RA}}} \leq \sum_{w=1}^{2h_N/q} N^{w - \lceil \frac{qw}{2} \rceil} g(w, N) h_N^{qw + \lfloor \frac{qw}{2} \rfloor - 2}$$
$$\leq \frac{2h_N}{q} \max_{1 \leq w \leq 2h_N/q} N^{w - \lceil \frac{qw}{2} \rceil} g(w, N) h_N^{qw + \lfloor \frac{qw}{2} \rfloor - 2}$$
$$\leq \frac{2}{q} N^{1 - \lceil \frac{q}{2} \rceil + \eta} g(1, N) h_N^{q + \lfloor \frac{q}{2} \rfloor - 1}$$
$$= O\left(N^{1 - \lceil q/2 \rceil + \eta}\right)$$

for large enough $N$ and for all $\eta > 0$. Note that for large enough $N$, $N^{w - \lceil \frac{qw}{2} \rceil}$ dominates $g(w, N) h_N^{qw + \lfloor \frac{qw}{2} \rfloor - 2}$, due to the conditions on $h_N$ stated in the lemma. Now, assume that the statement of the lemma is true for the case of $L - 1$.

We get

$$\sum_{h=1}^{h_N} \bar{s}_h^{\mathcal{C}_{\mathrm{RMA}(L)}}$$
$$= \sum_{w=\lceil \frac{q}{2^{L-1}} \rceil}^{2h_N} \bar{s}_w^{\mathcal{C}_{\mathrm{RMA}(L-1)}} \sum_{h=1}^{h_N} \sum_{d=1}^{\lfloor \frac{w}{2} \rfloor} \frac{\binom{N-h}{d}\binom{h-1}{d-1}\binom{h-d}{w-2d}}{\binom{N}{w}}$$
$$\leq \sum_{w=\lceil \frac{q}{2^{L-1}} \rceil}^{2h_N} \bar{s}_w^{\mathcal{C}_{\mathrm{RMA}(L-1)}} N^{-\lceil \frac{w}{2} \rceil} g'(w,N) \sum_{h=1}^{h_N} h^{w+\lfloor \frac{w}{2} \rfloor - 3}$$

where

$$g'(w,N) = (w)! \mathrm{e}^{w-1} \varphi_N(w-1) \sum_{d=1}^{\lfloor \frac{w}{2} \rfloor} \frac{(w-2d)^{2d-w}}{d^d(d-1)^{d-1}}$$

and $\mathcal{C}_{\mathrm{RMA}(l)}$ denotes the RMA code ensemble with $l$, $l = L-1, L$, accumulators. Note that the lower bound of $\lceil q/2^{L-1} \rceil$ in the summation over $w$ is due to the fact that the output size $h$ of an accumulator is at least $\lceil w/2 \rceil$, where $w$ is the input size. This is due to the binomial $\binom{h-d}{w-2d}$ and the upper bound of $\lfloor w/2 \rfloor$ in the summation over $d$. It follows that

$$\sum_{h=1}^{h_N} \bar{s}_h^{\mathcal{C}_{\mathrm{RMA}(L)}}$$
$$\leq \sum_{w=\lceil \frac{q}{2^{L-1}} \rceil}^{2h_N} \bar{s}_w^{\mathcal{C}_{\mathrm{RMA}(L-1)}} N^{-\lceil \frac{w}{2} \rceil} g'(w,N) h_N^{w+\lfloor \frac{w}{2} \rfloor - 2}$$
$$\leq O\left(N^{1-\sum_{i=1}^{L-1} \lceil \frac{q}{2^i} \rceil + \eta}\right)$$
$$\times \max_{\lceil \frac{q}{2^{L-1}} \rceil \leq w \leq 2h_N} N^{-\lceil \frac{w}{2} \rceil} g'(w,N) h_N^{w+\lfloor \frac{w}{2} \rfloor - 2}$$
$$= O\left(N^{1-\sum_{i=1}^{L} \lceil \frac{q}{2^i} \rceil + \eta}\right)$$

for large enough $N$ and for all $\eta > 0$. Above, we used the induction hypothesis in the second inequality. Also, note that for large enough $N$, $N^{-\lceil \frac{w}{2} \rceil}$ dominates $g'(w,N) h_N^{w+\lfloor \frac{w}{2} \rfloor - 2}$, due to the conditions on $h_N$ stated in the lemma. ∎

We can also extend [11, Lemma 27] to the stopping set case. For convenience, we state this lemma below.

*Lemma 6:* Let $r_{\mathrm{s}}^{\mathcal{C}_{\mathrm{RMA}}}(\rho; N)$ denote the $N$th stopping set size spectral shape function of the RMA code ensemble, defined as $r_{\mathrm{s}}^{\mathcal{C}_{\mathrm{RMA}}}(\rho; N) = \sup \frac{1}{N} \ln \bar{s}_{\lfloor \rho N \rfloor}^{\mathcal{C}_{\mathrm{RMA}}}$. Then,

$$r_{\mathrm{s}}^{\mathcal{C}_{\mathrm{RMA}}}(\rho; N) \leq \frac{2L \ln(N+1)}{N} + r_{\mathrm{s}}^{\mathcal{C}_{\mathrm{RMA}}}(\rho).$$

*Proof:* The proof of the lemma relies on the function $\psi(u, \rho)$, defined in (15). In particular, the proof of the lemma is by induction on $L$, following the same arguments as in the proof for the codeword case outlined in [12], and is therefore omitted for brevity. ∎

The final part of the proof of [11, Theorem 11] is also very general, and it can easily be extended to the stopping set case. In fact, the rest of the proof only relies on the following properties of $\psi(u, \rho)$.

- $\psi(u, \rho)$ is continuous;
- $\psi(u, \rho)$, for fixed $u$, is strictly increasing in $\rho < 1/2$;
- $\frac{\psi(u,\rho)}{u}$, for fixed $\rho$, is decreasing in $u$; and
- $\lim_{u \to 0} \frac{\psi(u,\rho)}{u} < 0 \ \forall \rho < \rho_0$.

The first three properties follow from (15), while the fourth property holds by assumption. Finally, by using Lemmas 5 and 6 and the properties above, Theorem 6 is proved following the same arguments as in the proof of [11, Theorem 28].

## ACKNOWLEDGMENT

The authors wish to thank Chiara Ravazzi, from Politecnico di Torino, for helpful discussions. The authors also wish to thank the anonymous reviewers and the Editor Rüdiger Urbanke for suggestions that improved the quality of the manuscript.